\documentclass[12pt, preprint]{aastex}

\shorttitle{Light Neutron-Capture Element Abundances in PNe I}
\shortauthors{Sterling et~al.}

\begin{document}

\title{The Abundances of Light Neutron-Capture Elements in Planetary Nebulae -- I.\ Photoionization Modeling and Ionization Corrections\protect \footnotemark \footnotetext{This paper includes data taken at the McDonald Observatory of the University of Texas at Austin.}}

\author{N. C. Sterling\altaffilmark{2, 3, 4}, Harriet L. Dinerstein\altaffilmark{2}, and T.\ R.\ Kallman\altaffilmark{4}}

\altaffiltext{2}{The University of Texas, Department of Astronomy, 1 University Station, C1400,
    Austin, TX 78712-0259; sterling@astro.as.utexas.edu, harriet@astro.as.utexas.edu}
\altaffiltext{3}{Presently a NASA Postdoctoral Fellow at the Goddard Space Flight Center.  The NASA Postdoctoral Program is administered by Oak Ridge Associated Universities through a contract with NASA.}
\altaffiltext{4}{NASA Goddard Space Flight Center, Code 662, Greenbelt, MD 20771; sterling@milkyway.gsfc.nasa.gov, tim@milkyway.gsfc.nasa.gov}

\begin{abstract}

We have conducted a large-scale survey of 120 planetary nebulae (PNe) to search for the near-infrared emission lines [\ion{Kr}{3}]~2.199 and [\ion{Se}{4}]~2.287~$\mu$m.  The neutron(\emph{n})-capture elements Se and Kr may be enriched in a PN if its progenitor star experienced \emph{s}-process nucleosynthesis and third dredge-up.  In order to determine Se and Kr abundances, we have added these elements to the atomic databases of the photoionization codes Cloudy and XSTAR, which we use to derive ionization correction factors (ICFs) to account for the abundances of unobserved Se and Kr ions.  However, much of the atomic data governing the ionization balance of these two elements are unknown, and have been approximated from general principles.  We find that uncertainties in the atomic data can lead to errors approaching 0.3~dex in the derived Se abundances, and up to 0.2--0.25~dex for Kr.  To reduce the uncertainties in the Kr ionization balance stemming from the approximate atomic data, we have modeled ten bright PNe in our sample, selected because they exhibit emission lines from multiple Kr ions in their optical and near-infrared spectra.  We have empirically adjusted the uncertain Kr atomic data until the observed line intensities of the various Kr ions are adequately reproduced by our models.  Using the adjusted Kr atomic data, we have computed a grid of models over a wide range of physical parameters (central star temperature, nebular density, and ionization parameter), and derived formulae that can be used to compute Se and Kr ICFs.  In the second paper of this series, we will apply these ICFs to our full sample of 120 PNe, which comprises the first large-scale survey of \emph{n}-capture elements in PNe.

\end{abstract}


\section{INTRODUCTION}

The compositions of planetary nebulae (PNe), the ejected envelopes of evolved low- and intermediate-mass stars (M~=~1--8~M$_{\odot}$), are affected by past nucleosynthesis in their progenitor stars.  In particular, these objects are important sources of C, N, and roughly half of the neutron(\emph{n})-capture element (atomic number $Z>30$) nuclei in the Universe (Busso et al.\ 1999, hereafter BGW99).

Carbon and \emph{n}-capture elements are produced during the thermally-pulsing asymptotic giant branch (AGB) phase.  Trans-iron nuclides are formed by slow \emph{n}-capture nucleosynthesis (the ``\emph{s}-process''), in which iron peak ``seed'' nuclei in the region between the H- and He-burning shells capture free neutrons released primarily by the reaction $^{13}$C($\alpha,n)^{16}$O (or, in more massive AGB stars, $^{22}$Ne($\alpha$,\emph{n})$^{25}$Mg).  The seed nuclei undergo a series of \emph{n}-captures and $\beta$-decays that transform them into heavier elements (K\"{a}ppeler et al.\ 1989; BGW99; Goriely \& Mowlavi 2000; Lugaro et al.\ 2003; Herwig 2005).

The intershell material, enriched in C (from partial He burning) and \emph{n}-capture elements, is conveyed to the stellar surface by convective mixing, a process denoted ``third dredge-up'' (TDU) to differentiate it from earlier dredge-up events during the red giant branch and early AGB phases.  TDU is believed to occur in stars with initial masses greater than 1.2--1.5~M$_{\odot}$, and is characterized by recurrent nuclear runaways of the He-burning shell (i.e., thermal pulses) separated by periods of convective mixing and transport to the surface (Iben \& Renzini 1983; BGW99; Herwig 2005; Straniero et al.\ 2006).  The enriched material is then released into the surrounding interstellar medium (ISM) by stellar winds and ultimately PN ejection.

Abundance determinations in PNe can be used to investigate the occurrence of the \emph{s}-process and TDU in PN progenitor stars, and their importance to the Galactic chemical evolution of \emph{n}-capture elements.  However, the low abundances of \emph{n}-capture elements ($\lesssim 10^{-9}$ relative to H; Asplund et al.\ 2005) cause their emission lines to be weak, and these elements were not seriously considered to be detectable in PNe until just over a decade ago.

It was not until 1994 that \emph{n}-capture element lines were identified in the spectrum of a PN (P\'{e}quignot \& Baluteau 1994, hereafter PB94).  These authors detected emission lines of Kr ($Z=36$), Xe ($Z=54$), and possibly other \emph{n}-capture elements in a deep optical spectrum of the bright PN NGC~7027.  Using approximations to the unknown Kr and Xe collision strengths (which have since been calculated by Sch\"{o}ning 1997 and Sch\"{o}ning \& Butler 1998) and corrections for the abundances of unobserved ionization stages, these authors found that Kr and Xe are highly enriched in NGC~7027, providing evidence for the occurrence of the \emph{s}-process and TDU in its progenitor star.

The study of PB94 led Dinerstein (2001) to realize that two long-unidentified emission lines seen in the $K$ band spectra of several PNe are in fact fine-structure transitions of [\ion{Kr}{3}] and [\ion{Se}{4}].  She found that the observed Kr and Se ($Z=34$) line strengths in IC~5117 and NGC~7027 are consistent with enhanced abundances, and furthermore proposed that the presence of these lines in some PNe but not others indicates a spread in \emph{s}-process enrichments among Galactic PNe.

Until now, the only other studies of \emph{n}-capture element abundances in PNe were those of Sterling et al.\ (2002) and Sterling \& Dinerstein (2003a), who detected Ge ($Z=32$) in absorption against the UV central star continua of six PNe with the \emph{Far Ultraviolet Spectroscopic Explorer} (\emph{FUSE}).  They derived Ge abundances in five of the objects, and found it to be enhanced in four PNe by factors of $\geq3$--10, depending on the amount of Ge depletion into dust.  For one of these objects, SwSt~1, Sterling et al.\ (2005) determined the line of sight Fe abundance from the \emph{FUSE} spectrum, and found it to be only slightly depleted ([Fe/S]~=~$-0.35\pm0.12$).  This implies that Ge is also negligibly depleted in SwSt~1, and that [Ge/S]~=~$0.72\pm0.06$, a factor of five enrichment relative to solar.

In all, these studies resulted in \emph{n}-capture element abundance determinations in only seven PNe.  Such a small sample of objects divulges little information regarding the \emph{s}-process and dredge-up histories of PN progenitors as a population, or their role in the Galactic chemical evolution of trans-iron species.

Therefore, we have undertaken a survey of Galactic PNe to search for the two near-infrared (NIR) lines identified by Dinerstein (2001), [\ion{Kr}{3}]~2.199 and [\ion{Se}{4}]~2.287~$\mu$m.  Se and Kr are useful tracers of \emph{s}-process enrichments and TDU in PN progenitor stars for two reasons: (1) Neither is expected to be depleted into dust, as Kr is a noble gas and Se has not been found to be significantly depleted in the diffuse ISM (Cardelli et al.\ 1993); and (2) the detection of the NIR lines of these elements in several PNe (e.g., Geballe et al.\ 1991; Hora et al.\ 1999; Lumsden et al.\ 2001) indicates that they may be observable in a significant fraction of Galactic PNe.  Furthermore, theoretical models predict that Se and Kr can be significantly enriched by \emph{s}-process nucleosynthesis in AGB stars (e.g., Gallino et al.\ 1998; Goriely \& Mowlavi 2000; Busso et al.\ 2001).  Therefore, if a star experiences the \emph{s}-process and efficient TDU, Se and Kr will be enriched in the PN it produces.

We have observed 103 PNe in the $K$ band, and utilize literature data to expand our sample to 120 objects.  This is the first large-scale survey of \emph{n}-capture elements in PNe, and comprises the first broad characterization of \emph{s}-process enrichments in PNe as a population.  We have detected Se and/or Kr in 81 of the 120 PNe in our sample, which increases the number of PNe with known \emph{n}-capture element abundances by more than a factor of ten.  Preliminary results from our survey have been presented in Sterling \& Dinerstein (2003b; 2004; 2005a,b; 2006) and Sterling (2006a).

In order to derive elemental Se and Kr abundances from our data, it is necessary to correct for the abundances of unobserved ionization stages.  We have added these elements to the photoionization codes Cloudy (Ferland et al.\ 1998) and XSTAR (Kallman \& Bautista 2001; Bautista \& Kallman 2001), which we use to derive widely applicable formulae for calculating Se and Kr ionization correction factors (ICFs).

This task is complicated by the general lack of atomic data governing the ionization balance of Se and Kr (e.g., photoionization cross-sections and rate coefficients for various recombination processes).  In fact, these data are unknown for almost \emph{all} \emph{n}-capture elements.  Therefore, we have used several approximations to calculate the atomic data of Se and Kr.  We have modeled ten PNe in our sample, which were selected because they exhibit emission lines from multiple Kr ions in their optical and NIR spectra.  This enables us to adjust the uncertain Kr atomic data until the observed line intensities of the various Kr ions are satisfactorily reproduced by our models.  Unfortunately, no other Se ions have been detected in PNe, and thus we are unable to empirically correct the uncertain Se atomic data.  With the adjusted Kr atomic data, we have run a grid of Cloudy models spanning a wide range of physical parameters (central star temperature, nebular density, and ionization parameter).  We extracted the fractional abundances of Se and Kr ions from this grid, and searched for correlations between these and the ionic fractions of more abundant elements.  The best correlations have been fit with analytical functions, which can be used to compute Se and Kr ICFs.  This paper, the first of two, describes these model calculations and presents the best analytical fits to the Se and Kr ICFs.  In the second paper of this series (Sterling \& Dinerstein 2007, hereafter Paper II), we will use our model-derived ICFs to determine Se and Kr abundances for our full sample of objects.

The outline of this paper is as follows: in \S2, we briefly describe our observations and supporting data from the literature for the 10 objects we model in detail.  In \S3, we present our models of these objects, empirically adjust the uncertain Kr atomic data, and discuss how uncertainties in the adopted atomic data affect the derived abundances.  In \S4, we present the grid of Cloudy models from which we derive formulae to correct for unobserved ionization stages of Se and Kr.  We discuss our results and provide concluding remarks in \S5.  Two appendices are also included.  Since the optical and NIR lines of Se and Kr have been frequently misidentified in the literature, in Appendix~A we investigate alternative identifications to the lines we consider in this study.  In particular, we remove important contaminants to the Se and Kr lines, and systematically rule out other possible identifications.  In Appendix~B, we describe the Se and Kr atomic data we have adopted in our photoionization models.

\section{OBSERVATIONS AND LITERATURE DATA}

\subsection{Infrared Spectra}

We have observed 103 PNe in the $K$ band with the CoolSpec spectrometer (Lester et al.\ 2000) on the 2.7-m Harlan J.\ Smith telescope at McDonald Observatory.  In addition, we utilize $K$ band PN spectra from the literature to expand our sample to 120 objects.  In Paper~II we will fully describe the observational data and reductions, and here we present only a brief synopsis.

Each PN was observed from 2.14--2.30~$\mu$m with a $2\farcs 7 \times90\arcsec$ slit and 75~l/mm grating, at a resolution of $R\sim$500 (FWHM~=~0.004~$\mu$m at 2.20~$\mu$m).  All data were reduced using IRAF\footnote{IRAF is distributed by the National Optical Astronomy Observatories, which are operated by the Association of Universities for Research in Astronomy, Inc., under cooperative agreement with the National Science Foundation.}.  For each two-dimensional spectrum, we performed dark subtraction, cosmic ray removal, and flat-fielding before extracting the one-dimensional spectrum.  The 1D spectra were then wavelength calibrated and dispersion-corrected using Ar lamp spectra.  For each PN, we observed at least one A0 standard star at a similar airmass, which we used to response-correct the spectrum, remove telluric features, and perform flux calibrations.

[\ion{Kr}{3}]~2.199 and [\ion{Se}{4}]~2.287~$\mu$m are resolved from nearby features of other species, except in PNe exhibiting vibrationally-excited molecular hydrogen.  In these objects, the Kr and Se lines may be contaminated by H$_2$~3-2~$S$(3)~2.201 and H$_2$~3-2~$S$(2)~2.287~$\mu$m, respectively.  As shown in Appendix~A.4, these lines are not important contaminants to the [\ion{Kr}{3}] and [\ion{Se}{4}] fluxes of the ten PNe considered in this paper.

\subsection{Optical Spectra and Source Selection}

We have selected 10 PNe from our sample that exhibit emission from [\ion{Kr}{4}] (and in one case, [\ion{Kr}{5}]) in their optical spectra, in addition to optical or NIR [\ion{Kr}{3}] emission lines.  The additional observed ionization stages reduce the uncertainties in correcting for unseen Kr ions, and enable us to empirically adjust the approximate Kr atomic data with the aid of photoionization models.  Unfortunately, no other Se ions have been clearly identified in PNe, and hence we are unable to optimize the Se atomic data.  (A possible exception is [\ion{Se}{3}]~$\lambda$8854.2, identified by PB94 in NGC~7027.  However, this line was undetected in the higher quality spectrum of Zhang et al.\ 2005, and hence we consider PB94's identification doubtful.)

The optical data references for these ten PNe are given in Table~1, along with the spectral resolution, extinction coefficients, and the sources of the NIR data and (when applicable) other useful information, such as elemental abundances and UV and IR line intensities.  The NIR line fluxes were dereddened using the extinction coefficients (Table~1) and adopted extinction law of the optical references.  We then applied an aperture correction to the NIR data (to account for the different slit widths and positions of the optical and NIR measurements) by forcing the \ion{H}{1} Br$\gamma$/H$\beta$ intensity ratios into agreement with the theoretical ratio at the $T_{\rm e}$ and $n_{\rm e}$ determined from the optical data, using Table~B.9 of Dopita \& Sutherland (2003).

The optical Kr line intensities are given in Table~2, where the cited 1-$\sigma$ errors are the flux uncertainties indicated in the optical data references.  Table~2 also lists the dereddened and aperture-corrected NIR Se and Kr line intensities of these objects.  The intensities have been corrected for contaminating blends as described in Appendix~A.

\section{PHOTOIONIZATION MODELS}

In order to derive Se and Kr abundances that are as accurate as possible, we have used two independent photoionization codes, Cloudy (Ferland et al.\ 1998) and XSTAR (Kallman \& Bautista 2001; Bautista \& Kallman 2001), to model the ionization structure of Se and Kr.  We have expanded the atomic database for each of these codes to include Se and Kr, and performed detailed spectral fits of the ten PNe listed in Table~1.

The Se and Kr atomic data that we have utilized in Cloudy and XSTAR are discussed in Appendix~B.  Much of the atomic data controlling the ionization balance of these elements is unknown, in particular the photoionization (PI) cross-sections and rate coefficients for radiative recombination (RR), dielectronic recombination (DR), and charge transfer (CT).  As explained in Appendix~B, we have used various approximations to calculate values for the cross-sections and rate coefficients of these atomic processes.

The atomic data has been entered into Cloudy version C06.01 and XSTAR version 2.1kn3.  Cloudy loops through all elements up to the maximum atomic number in its database, and therefore it was necessary to enter atomic data for the remaining elements between Zn (the heaviest element in vC06.01) and Kr.  We computed PI cross-sections and RR rate coefficients for Ga, Ge, As, and Br using the same methods described in Appendix~B, but assumed the CT, DR, and collisional ionization rate coefficients to be the same as the corresponding charge states of Se or Kr.  These four elements were turned off in all of our Cloudy models, and therefore do not affect our results.  In contrast, it was not necessary to enter data for these intervening elements in XSTAR.

Se and Kr are trace elements in PNe, even when enriched by the \emph{s}-process.  As such, they are not expected to significantly affect the thermal or ionization structure and predicted spectra of nebulae.  For each of the models described in this section, we also ran the corresponding unaltered versions of Cloudy and XSTAR (without Se and Kr) with the same input parameters to test this assumption.  In all cases, we found negligible differences between the models with and without Se and Kr, indicating that our additions to the atomic databases did not introduce any unforeseen effects in the functionality of these codes.

In the following, we present detailed models of ten PNe in which the input parameters were optimized to best reproduce the observed UV, optical, and IR spectra.  We discuss uncertainties in the derived Se and Kr abundances in these nebulae arising from a variety of sources, including the atomic data, and how these will affect the derived abundances of our full sample of 120 PNe.

\subsection{Cloudy Models}

We performed detailed spectral fits to the UV, optical, and IR line intensities of the ten PNe in Table~1, in order to reproduce their ionization structure.  The main objective of this exercise was to compare the predicted Kr line strengths to the observed ones, and to make necessary adjustments to the uncertain Kr atomic data (Appendix~B) so that these lines are best reproduced by our models.  Once agreement was achieved, we incorporated the optimized Kr atomic data into our grid of Cloudy models (\S4), which we use to derive ICFs for Se and Kr.

We made a number of simplifying assumptions in these models.  First, spherical geometry was assumed, with filling and covering factors (the fraction of material in a solid angle of 4$\pi$ steradians around the central star) of unity.  Second, we assumed the nebulae have constant densities.  Since this produced reasonably good fits to the observed spectra, we did not attempt to use other density laws.  Finally, we do not include dust physics in our models of single objects, aside from the depletion of refractory elements into grains.  This is likely to result in an inaccurate treatment of the thermal balance of the nebulae, due to the influence of photoelectric heating by dust (Dopita \& Sutherland 2000).  For this reason, the exclusion of dust may also affect the derived central star temperatures.  In \S4, we show that ignoring dust physics has a negligible effect on the predicted ionization balance of Se and Kr.

These simplifications are not likely to be important for the purposes of our study.  Our goal is not to produce physically accurate models of these PNe, but rather to establish the ionization balance of the nebulae in order to adjust the approximate Kr atomic data.  For this reason, our model-derived abundances for elements other than Se and Kr should not be regarded as improvements over those of the references listed in Table~1.

For the input parameters of each model, we initially adopted the abundances and densities reported in the literature (see Table~1).  It is also necessary to specify the stellar flux distribution, inner and outer radii of the nebular material, and ionizing luminosity of the central star.  

In all cases, we utilized NLTE stellar atmospheres including metal line blanketing, computed for hot compact stars such as the central stars of PNe (Rauch 2003).  Solar abundances were assumed for the model atmospheres, since the modeled PNe are all Galactic disk objects likely to have near-solar abundances.  Since the Rauch (2003) grid only extends from 50,000~K to 190,000~K, we were unable to use it to model IC~418, which has a very cool central star.  T.\ Rauch (2005, private communication) kindly computed models with $T_{\rm eff}=30,000$ and 40,000~K and log($g$)~=~3.4, which we use to model IC~418.  The Rauch stellar atmospheres do not include mass-loss, which can significantly alter the ionizing flux of a PN central star (Kudritzki \& Pols 2000).  We have tested the dependence of our results on this factor by modeling IC~418 with a stellar atmosphere including mass-loss (Koesterke et al.\ 2007, in preparation), and find that incorporating mass-loss does not significantly affect the derived nebular abundances (\S3.1.1).

For the nebular radii and stellar luminosities, we used as initial parameters the quantities derived from the photoionization models of the Hyung and Aller references cited in Table~1, in addition to Hyung et al.\ (1994b) for IC~418, Hyung \& Aller (1997a) for NGC~6741, Hyung et al.\ (1997) for NGC~6884, Keyes et al.\ (1990) for NGC~7027, and Hyung \& Aller (1997b) for NGC~7662.  We also adopted the stellar gravities of these references for the Rauch NLTE models, except for IC~418, where we used log($g$)~$=3.4$ rather than the value of 2.9 computed by Hyung et al.\ (note that M\'{e}ndez et al.\ 1992 found log($g)=3.45$ for this star).  In the case of NGC~6826, we used the nebular radius, $T_{\rm eff}$, and stellar luminosity of M\'{e}ndez et al.\ (1992) as initial parameters.  Instead of log($g$)~=~4.0, as computed by those authors, we used log($g$)~=~5.0, which is the minimum gravity in the grid of Rauch (2003) NLTE atmospheres.

With initial values for all nebular parameters in hand, we varied the abundances, $T_{\rm eff}$, $n_{\rm H}$, and in many cases the outer radii and stellar luminosity until a good fit between the predicted and observed spectra was obtained.  We used a mixture of the subplex optimization routine in Cloudy and manual manipulation of the parameters to achieve these fits.  The input parameters for all our models are given in Tables~3 and 4 (with the solar composition in the last column for comparison).  In Tables~5, 6, and 7, we display the predicted versus observed line intensities (relative to H$\beta$) for lines used in the spectral fits.  The intensities are all on the scale $I$(H$\beta$)~=~100.  For most lines, the predicted intensities are within 25\% of the observed ones.  In some cases, there are larger discrepancies; these are likely due to the simplifying assumptions discussed above.

With the Kr atomic data we originally calculated, we found that the predicted [\ion{Kr}{3}] lines were too strong, and the [\ion{Kr}{4}] lines too weak relative to the observed intensities.  In addition, the [\ion{Kr}{5}]~$\lambda$6256.1 intensity was underproduced in NGC~7027, the only PN of our sample in which this line was clearly detected.  Increasing the Kr$^{++}$ PI cross-section derived from Equation (B1) by a factor of two, and decreasing that of Kr$^{3+}$ by a factor of four leads to better agreement in the relative line intensities of [\ion{Kr}{3}], [\ion{Kr}{4}], and [\ion{Kr}{5}].

We adopt these optimized PI cross-sections (listed in Table~B.2) in our models.  The predicted Kr line strengths in Tables~5--7 are from models using the adjusted Kr$^{++}$ and Kr$^{3+}$ PI cross-sections.  As seen in these tables, our models do not predict the Kr ionization balance to be systematically too high or low.  The differences between the predicted and observed Kr line intensities can be attributed to flux uncertainties ($\sim$30--40\%) for all but two objects.  In NGC~6884, [\ion{Kr}{3}]~$\lambda6826.7$ is predicted to be too strong by 60\%, but as discussed in Appendix~A.3, the flux of this line is uncertain due to possible blending with \ion{He}{1} and telluric OH features.  In IC~418, [\ion{Kr}{3}]~2.199~$\mu$m is predicted to be too strong, while [\ion{Kr}{4}] lines are too weak.  We were unable to reproduce the observed [\ion{Kr}{4}] line intensities in this object except with a higher $T_{\rm eff}$, which led to a poor fit to the rest of the spectrum.

Note that the derived Se and Kr abundances given in Tables~3--4 are often higher than the solar Se and Kr abundances (3.36$\pm$0.04 and 3.28$\pm$0.08, respectively) of Asplund et al.\ (2005), particularly in the case of Kr.  In many cases, these enrichments are too large to be explained by the observed star-to-star scatter in the initial abundances of \emph{n}-capture elements in stars of near-solar metallicity (Travaglio et al.\ 2004; Simmerer et al.\ 2004).  Such large abundance enhancements must instead be due to self-enrichment by \emph{s}-process nucleosynthesis (BGW99).  In addition, theoretical \emph{s}-process models predict Kr to be more enriched than Se by 0.0--0.5 dex (Gallino et al.\ 1998; Goriely \& Mowlavi 2000; Busso et al.\ 2001), as we have found for most of these PNe.  We believe that the subsolar Se abundances we derive for NGC~6790 and NGC~6826 are likely to be reflections of the overall lower metallicities of these objects rather than any type of Se depletion (e.g., into dust), based on the subsolar O, Ne, and Ar abundances determined for these objects by Liu et al.\ (2004b).  We reserve a full discussion of the \emph{s}-process enrichments in these objects for Paper~II.

\subsubsection{Dependence of IC~418 Model on Stellar Mass-Loss}

In order to test the sensitivity of our results to the effects of stellar winds on the ionizing fluxes of the central stars (Kudritzki \& Pols 2000), we have modeled IC~418 with a stellar atmosphere including mass-loss, computed for us by L.\ Koesterke (2005, private communication; see also Koesterke et al.\ 2007, in preparation).  This stellar wind model was computed by fitting a synthetic spectrum to \emph{International Ultraviolet Explorer} (\emph{IUE}) low-dispersion SWP spectrum retrieved from the Multi-Mission Archive at the Space Telescope Science Institute (MAST).  The low-dispersion SWP datasets taken with the large aperture (SWP03177, SWP03178, SWP03810, SWP06651, SWP43170, SWP43171, SWP46377, and SWP46378), were weighted by exposure time and co-added to increase the signal-to-noise ratio.  A Rauch NLTE atmosphere with no mass-loss was used initially, and the mass-loss rate $\dot{M}$ was increased until the resonance line P~Cygni profiles in the \emph{IUE} spectrum were reproduced.  Two stellar models were computed, with $T_{\rm eff}$~=~35,000 and 40,000~K, log($g$)~=~3.4, and $\dot{M}=10^{-6}$~M$_{\odot}$~yr$^{-1}$, which we interpolated to produce a Cloudy model of IC~418.

Our best-fit values for the stellar and nebular properties using the Koesterke stellar atmosphere are quite similar to those with the Rauch (2003) atmosphere (see Tables~3 and 5), with the $T_{\rm eff}$ agreeing to within 5\% and the nebular abundances to within 30\% with the exception of O, Ne, and Kr.  The reason for these discrepant abundances lies in the flux distributions of the two stellar atmospheres.  The mass-losing atmosphere of Koesterke emits more O$^{+}$ ionizing photons, and therefore produces O$^{++}$ more easily than the Rauch (2003) model.  This results in a better fit to the observed O lines, and it is not necessary to increase the O abundance to produce agreement with the observed [\ion{O}{3}] line intensities, as is necessary with the Rauch atmosphere.  Similarly, the wind atmosphere ionizes Kr$^{++}$ more easily than the Rauch atmosphere, allowing the [\ion{Kr}{4}] lines to be fit with a lower Kr abundance.  In contrast, the Koesterke atmosphere does not produce enough Ne$^{+}$ ionizing photons, which does not permit a good fit to the Ne line intensities.  It was necessary to increase the Ne abundance until the $\chi^2$ of the [\ion{Ne}{2}] and [\ion{Ne}{3}] line fits was minimized (this does not significantly affect the Kr ionization balance).

The only other modeled PNe which exhibit wind signatures in their \emph{IUE} spectra are NGC~6572 and NGC~6826.  Given the general agreement between Cloudy models of IC~418 using stellar atmospheres with and without mass-loss, we have not attempted to model these other objects with atmospheres including stellar winds.

\subsubsection{Error Analysis}

We have estimated uncertainties for the derived $T_{\rm eff}$ and the Se and Kr abundances of each modeled PN.  We do not present formal error bars for the abundances of other elements, since it was not our objective to determine accurate abundances for those species.

Our error estimates for Se and Kr include uncertainties in the line intensities, $T_{\rm eff}$, and hydrogen density $n_{\rm H}$.  First, we estimated the uncertainties in the adopted $T_{\rm eff}$ by manually adjusting this parameter in the models until the nebular temperature diagnostics and/or ionization balance were no longer acceptably reproduced (to within 50--100\%).  These uncertainties are given in Tables~3--4.  We then re-derived the Se and Kr abundances with the maximum and minimum allowed $T_{\rm eff}$, to determine their sensitivity to uncertainties in the assumed $T_{\rm eff}$.  Next, we computed the Se and Kr abundances after changing the density by a factor of two (also adjusting the outer radii of the nebulae so that the models did not become too optically thin or thick).  Finally, we estimated the errors due to uncertainties in the Se and Kr line intensities.  For Kr, this involved adjusting the abundance until the most discrepant line intensities in Tables~5--7 were reproduced; if all Kr lines were predicted with good accuracy (e.g., NGC~7027), we simply assumed errors of 30\% in the line strengths.  On the other hand, only one line of Se was observed for each object.  The flux errors for this line (Table~2) lead to uncertainties of only 5--25\% in the Se abundance, likely resulting in underestimated formal errors to the Se abundances.

In Table~8, we compare the Se and Kr abundance errors (in dex) from each source of uncertainty we considered.  In general, the line strengths are the greatest source of uncertainty for the Kr abundances.  However, the errors from $T_{\rm eff}$ uncertainties are large in the low excitation PNe IC~418 and NGC~6826, where the Kr ionization balance is strongly sensitive to the stellar temperature.  For Se, the abundance errors have comparable contributions from uncertainties in the line strengths, $T_{\rm eff}$, and $n_{\rm H}$.  We added the errors in quadrature to obtain the formal (1-$\sigma$) abundance uncertainties given in Tables~3, 4, and 8.

\subsubsection{Abundance Errors From Atomic Data Uncertainties}

One of the most important sources of uncertainty in our derived Se and Kr abundances is the approximate atomic data we have adopted.  In order to gain insight into the magnitude of the abundance uncertainties introduced by the use of the approximate atomic data, we have run Monte Carlo simulations for four of our Cloudy models, in which the threshold PI cross-sections and rate coefficients of RR, CT, and low-$T$ DR have been varied separately.  The Monte Carlo simulations were run by using a Gaussian random number generator included in the Cloudy distribution.

As described in Appendix~B, we have used the Kramers (1923) formula to derive threshold PI cross-sections for Se and Kr ions.  Gould (1978) showed that this approximation is accurate to within a factor of two for first and second row elements of the Periodic Table.  Since Se and Kr are significantly more complex systems, the Kramers formula may not be as accurate.  Therefore, we have varied the Se and Kr PI cross-sections in Cloudy by allowing each threshold cross-section to randomly take values within a Gaussian distribution of FWHM 0.5~dex (a factor of three uncertainty, 1-$\sigma$) around the value given in Table~B2.  We did not attempt to investigate deviations from the adopted \emph{shape} of the cross-section as a function of energy (a power law of index $-2$), and this remains an important uncertainty.  We varied the rate coefficients for CT, low-$T$ DR, and RR by the same amount.  While the RR rate coefficients were derived in a self-consistent manner using the Milne relation and the PI cross-sections, the CT and low-$T$ DR rate coefficients could not be estimated from first principles.  Instead, we assumed that Se and Kr ions have the same rate coefficients for these processes as similar charge states of nearby elements.  The 0.5~dex 1-$\sigma$ uncertainties we adopt for the CT and low-$T$ DR rate coefficients are based on the expected uncertainties of Landau-Zener CT calculations (Kingdon \& Ferland 1996) and the error bars of the low-$T$ DR rate coefficient estimates in Cloudy (Ferland et al.\ 1998), but may underestimate the actual uncertainties for Se and Kr ions.

We chose four PNe to illustrate the errors from atomic data uncertainties: IC~418, NGC~6741, NGC~6884, and NGC~7027.  These objects were selected because they span a large range of nebular density and excitation.  For each of these PNe we ran 100 Cloudy models, in which the PI cross-sections and rate coefficients for RR, CT, and low-$T$ DR were separately varied for all Se and Kr ions.  The atomic data of each ion was varied independently of other ions.

In Table~9, we display the results of our Monte Carlo simulations, showing the errors (in dex) of predicted line intensities for each observed ion.  The intensity errors reflect uncertainties in the abundance of the parent ion, which in turn affects the derived elemental abundances.  It can be seen that errors in the PI cross-sections cause the greatest uncertainties in the predicted line intensities, followed by the CT, low-$T$ DR, and RR rate coefficients.  Taken together, the atomic data uncertainties alone can result in errors of nearly a factor of two (0.3~dex) in the derived Se abundances.  In the case of Kr, if only [\ion{Kr}{3}] is observed (as for most objects in our full sample), the abundance errors can be as large as 0.26~dex from the atomic data uncertainties.  However, since we have empirically adjusted the Kr atomic data, the abundances are not likely to be so sensitive to atomic data uncertainties.

This exercise reveals the sensitivity of our analysis to the unknown Se and Kr atomic data.  Without a dedicated effort to calculate and experimentally measure these atomic data, uncertainties in the ionization balance for Se and Kr will not allow their abundances to be derived to accuracies of better than 0.2--0.3~dex.  In fact, when other sources of error (e.g., observational uncertainties) are taken into account, their abundances may be considerably more uncertain.

\subsection{XSTAR Models}

In order to test the efficacy of the Se and Kr ionization balance and abundance results of our Cloudy models, we have modeled the ten objects listed in Table~1 with the photoionization code XSTAR (Kallman \& Bautista 2001; Bautista \& Kallman 2001).  This code provides an independent test of our Cloudy results.

XSTAR is a 1D photoionization code, like Cloudy, but does not incorporate dust physics.  Therefore, we used the same model assumptions (spherical geometry, constant density, no dust) as for the Cloudy models.  Since XSTAR models are significantly more computationally demanding than Cloudy, we did not attempt to fit the observed spectra by optimizing the input parameters.  Instead, we used the best-fit Cloudy values as input parameters to these models, although it occasionally was necessary to adjust the outer radii slightly so that the models were not too optically thick or thin.

We find that the XSTAR models reproduce the predicted Cloudy spectrum for each PN reasonably well in all cases, with the intensities agreeing to within 30\% for most lines.  In particular, the [\ion{Se}{4}]~2.287~$\mu$m line intensities agree with the Cloudy results to within 30\%, and often better, for all of the modeled PNe.  The predicted Kr line intensities show larger differences from the Cloudy results, sometimes by as much as 50\%.  However, no systematic trends were apparent for the models with discrepant Kr line strengths.  We therefore conclude that the Se and Kr abundance errors introduced by using different photoionization codes ($\leq$30--50\%) are less than those of other sources of uncertainty (line intensities, $T_{\rm eff}$, densities, and atomic data errors).  Our ionization balance and abundance results for Se and Kr thus are not strongly model-dependent.

\section{IONIZATION CORRECTION FACTORS FOR SE AND KR}

The primary goal of this study is to derive widely applicable recipes to correct for the abundances of unobserved ionization stages of Se and Kr, so that their elemental abundances may be determined.  For the vast majority of the PNe in our full sample (120 objects), we have observed only one ionization stage of each of these elements: Se$^{3+}$ and Kr$^{++}$.

Therefore, we have constructed a grid of Cloudy models spanning a large range of physical parameter space, to search for correlations between the fractional abundances of Se$^{3+}$ and Kr$^{++}$ and ionic fractions of more commonly detected elements.  In particular, the grid includes models for various values of $T_{\rm eff}$, $n_{\rm H}$, and ionization parameter $U=Q$(H)$/(4\pi R_{\rm in}^2 n_{\rm H} c)$, where $Q$(H) is the hydrogen-ionizing photon density, $R_{\rm in}$ the inner radius of the nebula, and $c$ the speed of light.  Given the density, the ionization parameter essentially determines the ionizing luminosity of the central star.

We list in Table~10 the full set of physical parameters which were varied in our grid of models.  A model was run for every combination of $T_{\rm eff}$, $n_{\rm H}$, and log($U$) in Table~10, for a total of 3761 models.

In our grid of models we have ignored dust, and assumed the default PN abundances in Cloudy (from Aller \& Czyzak 1983 and Khromov 1989).  We have tested the sensitivity of our results to metallicity and the presence of dust, which we discuss in the subsections below.  All models have spherical geometry, constant density, an inner radius of $10^{-2.5}$~pc, and extend to an outer radius where $T_{\rm e}$ falls below 4000~K.  Rauch (2003) NLTE stellar atmospheres with log($g$)~=~6.0 and solar abundances were used in all models.

For each model, we extracted the fractional abundances (averaged over volume) of all ions of He, C, N, O, and Ne, and the first 10 ions of S, Cl, Ar, Se, and Kr, and compared the Se and Kr fractional abundances with several ionic fractions (or combinations thereof) of these other elements.  In Figure~1, we plot the best correlations between the observed Se and Kr ions and other ionic ratios.  We also include correlations with (Kr$^{++}+$~Kr$^{3+}$)/Kr, for completeness.  In the six panels of Figure~1, the fractional abundances from each model in our grid are represented as dots.  In some of the panels, e.g., Se$^{3+}$/Se vs.\ O$^{++}$/O, a family of curves is seen among the plotted ionic fractions.  Each curve corresponds to a single value of $T_{\rm eff}$, indicating the correlation's sensitivity to the nebular gas temperature (which generally increases with $T_{\rm eff}$).

We have performed fits to the ionic fraction correlations, using a $\chi^2$ minimization routine written in IDL, and considering polynomial and exponential functional forms.  These are displayed in Figure~1 as solid lines.  The inverses of these fitting functions correspond to ICF formulae which can be used to convert Se and Kr ionic abundances into elemental abundances.

In particular, when [\ion{Kr}{3}] is the only Kr ion observed, the Kr ICF may be derived using the fractional abundance of either S$^{++}$ or Ar$^{++}$, which are commonly detected in optical spectra of PNe.  The ICFs are:
\begin{equation}
\mathrm{ICF}(\mathrm{Kr}) = \mathrm{Kr}/ \mathrm{Kr}^{++} = (-0.009205 + 0.3098x + 0.0007978e^{6.297x})^{-1},
\end{equation}
\begin{displaymath}
x = \mathrm{Ar}^{++}/\mathrm{Ar} \geq 0.027;
\end{displaymath}
and
\begin{equation}
\mathrm{ICF}(\mathrm{Kr}) = \mathrm{Kr}/ \mathrm{Kr}^{++} = (-0.3817 + 0.3796e^{1.083y})^{-1},
\end{equation}
\begin{displaymath}
y = \mathrm{S}^{++}/\mathrm{S} \geq 0.0051.
\end{displaymath}
The lower limits to $x$ and $y$ denote the range of validity for these ICFs (below these limits, the ICFs become negative).  Of these two ICFs, we recommend Equation~(1) over Equation~(2) since more ions of Ar (Ar$^{++}$, Ar$^{3+}$, and Ar$^{4+}$) are typically observed in optical spectra of PNe than of S (S$^{+}$ and S$^{++}$).  This leads to a more accurate Ar abundance from optical data than that of S.  In fact, Henry et al.\ (2004) have found that S abundances in PNe may be inaccurate when determined from optical data and model-derived ICFs, and explain this behavior by the inability of photoionization models to accurately compute the abundance of S$^{3+}$, which can only be observed with IR spectroscopy.

Nevertheless, both correlations (top panels of Figure~1) are quite strong, and if the Ar and S abundances are well-determined, both should be reliable if our models accurately reproduce the Kr ionization equilibrium.  In Paper~II, we make use of both of these equations.

On the other hand, correlations between Se$^{3+}$/Se and ionic fractions of other elements are not as well-defined.  This will lead to less accurate ICFs, and hence less accurate Se abundance determinations.  In the middle left panel of Figure~1, we display the fit to the best correlation we found, between Se$^{3+}$/Se and O$^{++}$/O.  This corresponds to
\begin{equation}
\mathrm{ICF}(\mathrm{Se}) = \mathrm{Se}/ \mathrm{Se}^{3+} = (-0.1572 - 0.3532z^{17.56} + 0.153e^{1.666z})^{-1},
\end{equation}
\begin{displaymath}
z = \mathrm{O}^{++}/\mathrm{O} \geq 0.01626.
\end{displaymath}

We have also derived ICFs to be used when both [\ion{Kr}{3}] and [\ion{Kr}{4}] are observed.  Although we do not utilize these ICFs in our study, we present them for completeness.  The ICF that is least sensitive to variations in $n_{\rm e}$, $T_{\rm eff}$, and $U$ is
\begin{equation}
\mathrm{ICF}(\mathrm{Kr}) = \mathrm{Kr}/ (\mathrm{Kr}^{++} + \mathrm{Kr}^{3+}) = (-0.04996 + 0.7041x + 0.3679x^{2})^{-1},
\end{equation}
\begin{displaymath}
x = (\mathrm{Cl}^{++} + \mathrm{Cl}^{3+}) / \mathrm{Cl} \geq 0.0686.
\end{displaymath}
However, in order to detect [\ion{Cl}{4}], it is necessary to have obtained a spectrum extending to at least 7550~\AA.  For data which do not extend to such long wavelengths, we offer the following ICFs:
\begin{equation}
\mathrm{ICF}(\mathrm{Kr}) = \mathrm{Kr}/ (\mathrm{Kr}^{++} + \mathrm{Kr}^{3+}) = (-0.1549 + 0.6458y + 0.0048e^{4.657y})^{-1},
\end{equation}
\begin{displaymath}
y = (\mathrm{Ar}^{++} + \mathrm{Ar}^{3+}) / \mathrm{Ar} \geq 0.2193;
\end{displaymath}
\begin{equation}
\mathrm{ICF}(\mathrm{Kr}) = \mathrm{Kr}/ (\mathrm{Kr}^{++} + \mathrm{Kr}^{3+}) = (-0.2074 + 0.9834z + 0.000513e^{5.983z})^{-1},
\end{equation}
\begin{displaymath}
z = (\mathrm{O}^{+} + \mathrm{O}^{++}) / \mathrm{O} \geq 0.21.
\end{displaymath}
Both of these correlations exhibit large dispersion and a non-trivial dependence on $T_{\rm eff}$ and log($U$), and are less accurate than Equation~(4) unless $y$ and $z$ are large (the dispersion about the fit becomes small in this case; see the lower panels of Figure~1).

\subsection{Dependence of ICFs on Metallicity}

To derive Equations (1)--(6), we have used the default PN abundances of Cloudy.  However, some of the PNe of our sample may have substantially different abundances.  In order to test the robustness of our ICFs to the assumed metallicity, we have run the same grid of models for low metallicity PNe by dividing the default Cloudy PN abundances by 10 for all elements heavier than He, and using NLTE stellar atmospheres with halo abundances (Rauch 2003).  Note that these abundances were not chosen to represent specific PNe, but are simply used to illustrate that our derived ICFs are not strongly dependent on the adopted nebular abundances.

In Figure~2, we show the Se and Kr ionic fraction correlations discussed above when low metallicity abundances are adopted.  We overplot the fits to the correlations of Figure~1 as solid lines.  As can be seen from Figure~2, adopting lower metal abundances does not significantly affect the correlations we found in the previous section.  In particular, the correlations described by Equations~(1)--(4) for high metallicities are remarkably similar to those at low metallicities (although slightly less Se$^{3+}$ is produced at high O$^{++}$ fractions in the low metallicity models).  The only significant discrepancies in the correlations are for those in the lower panels of Figures~1 and 2, where we find higher (Kr$^{++}+$~Kr$^{3+}$) fractional abundances in our low metallicity models for low (Ar$^{++}+$~Ar$^{3+}$)/Ar and (O$^{+}+$~O$^{++}$)/O.  This signifies a deterioration in the accuracy of Equations~(5) and (6) for low metallicity, low excitation PNe.  However, these correlations exhibit a large dispersion even at higher metallicities, and are not as reliable as the correlation of Equation~(4), which applies to low metallicity PNe.  We conclude that adopting a lower metallicity does not significantly affect the ICFs of Equations~(1)--(4).

\subsection{Dependence of ICFs on Dust}

To test the dependence of our ICFs on the inclusion of dust physics, we have run our grid of models with C-rich grains (the ``grain Orion graphite'' command in Cloudy) and O-rich dust (the ``grain Orion silicate'' command), with grain properties from Baldwin et al.\ (1991) and van~Hoof et al.\ (2004).

In Figure~3, we display the ionic fraction correlations for Cloudy models including C-rich dust.  While some minor differences in the fractional abundances of these ions are apparent when compared to Figure~1, the ICFs derived from models without dust (overplotted as solid lines) also apply to models that include dust.

In our O-rich dust models, the silicate grains began sublimating for high values of log($U$) ($\geq 1.0$), and hence we were not able to run our full grid.  However, the correlation plots (not shown) are similar to those of the C-rich dust models, and again no significant differences are seen compared to the dust-free models.  We therefore conclude that the presence of dust does not affect our derived ICFs.

\subsection{Comparison of ICF- and Model-Derived Se and Kr Abundances}

In this section, we compare the Se and Kr abundances determined from our ICFs to those found from our models of the ten PNe listed in Table~1 (\S3.1).  Since our interest is primarily in determining Se and Kr abundances from their NIR lines, we only consider the ICFs given by Equations (1)--(3).

For each of the ten PNe we have modeled, we have taken ionic and elemental abundances needed for our ICFs (Ar$^{++}$/Ar, S$^{++}$/S, and O$^{++}$/O) from the literature (see Table~1).  To properly account for uncertainties in the derived ICFs, we consider errors in these ionic and elemental abundances.  Unfortunately, such an error analysis has not been performed in any of the abundance references we use.  Therefore, we adopt 20\% uncertainties in the O, S, and Ar ionic and elemental abundances (see Paper~II).  We propagate these errors into our ICFs, as well as the standard deviations of our fits to the ionic fraction correlations in Figure~1.

To determine the Kr$^{++}$ and Se$^{3+}$ ionic abundances, we have utilized 5-level and 2-level model atoms, respectively, with transition probabilities and effective collision strengths from the references listed in Table~B1.  In these calculations, we have used the $T_{\rm e}$([\ion{O}{3}]) and $n_{\rm e}$ derived in the optical data references.  Uncertainties of $\pm$1000~K in $T_{\rm e}$ and 20\% in $n_{\rm e}$ were assigned when not reported in these references, and propagated into the Se$^{3+}$ and Kr$^{++}$ abundance errors.

We present the results of our calculations for Kr in Table~11 and Se in Table~12.  In these tables, the ICFs derived from Equations (1)--(3) are given, followed by the abundances determined using our ICFs and photoionization models.  Despite the considerable uncertainties we estimate for the ICFs, we find gratifying agreement between the ICF and model abundance determinations.  In all cases, the abundances agree within the errors, and with a couple of exceptions agree to within 0.15 dex ($\sim$40\%) for Kr and 0.1 dex ($\sim$25\%) for Se.  This also illustrates that the simplifying assumptions we made in our models do not significantly affect the derived Se and Kr abundances.

The largest discrepancies in the Kr abundances occur for objects in which only \linebreak\ [\ion{Kr}{3}]~$\lambda$6826.7 was detected, while 2.199~$\mu$m was unobserved.  As discussed in Appendix~A.3, $\lambda$6826.7 can be blended with \ion{He}{1}~$\lambda$6827.9 and telluric OH emission in low resolution spectra.  Therefore, the intensity of [\ion{Kr}{3}]~$\lambda$6826.7 may be quite uncertain if these contaminants are not properly removed.

Another major source of error in our ICF-derived abundances stems from the ICF itself.  In some cases, the uncertainties in the ICF approach a factor of two.  Furthermore, for small O$^{++}$ or Ar$^{++}$ fractions, the ICF becomes large, and may conceivably be more uncertain than we have estimated.  We believe that the uncertainties in the ICFs and [\ion{Kr}{3}]~$\lambda$6826.7 intensities are able to explain the largest discrepancies between our ICF- and model-derived Se and Kr abundances, namely for NGC~6741, NGC~6826, and NGC~7662.  Nevertheless, the overall agreement between the Se and Kr abundances derived with these different methods is encouraging, and validates the ICFs we have derived.

\section{CONCLUDING REMARKS}

We have presented results of a photoionization model study designed to derive correction factors for the abundances of unobserved Se and Kr ions.  These corrections are needed to determine elemental abundances from our survey of [\ion{Kr}{3}]~2.199 and [\ion{Se}{4}]~2.287~$\mu$m in 120 Galactic PNe, the first large-scale survey of \emph{n}-capture elements in PNe.  We have added Se and Kr to the atomic databases of the photoionization codes Cloudy and XSTAR.  While the atomic data concerning energy transitions for the observed Se and Kr ions are known (i.e., transition probabilities and effective collision strengths), much of the data governing the ionization equilibrium of these elements have not been determined.  We have approximated the rate coefficients for these processes as outlined in Appendix~B.

When deriving ICFs, it is critical to accurately solve for the ionization balance of the element in question.  To this end, we calculated detailed models of ten PNe exhibiting emission lines from multiple ionization states of Kr.  The Kr atomic data were then empirically adjusted until the emission line intensities from the various Kr ions were adequately reproduced by our models.  Unfortunately, no other Se ions have been clearly detected in PNe, and thus we could not empirically correct the uncertain Se atomic data we adopted.  This makes the Se abundances we derive from our survey inherently more uncertain than those of Kr.

Using the adjusted Kr atomic data, we computed a grid of Cloudy models spanning a large range of $T_{\rm eff}$, $n_{\rm H}$, and ionization parameter $U$, which encompass the values of these parameters likely to be encountered in most PNe.  From these grids, we extracted the fractional abundances of the observed Se and Kr ions, and compared them to the ionic fractions of commonly detected elements.  We computed fits to the best correlations we found, which serve as ICFs that can be used to determine Se and Kr elemental abundances when only one or two ions of these species are observed.  The ICF formulae are given in Equations~(1)--(6).

We have compared the Se and Kr abundances derived with the ICF method with those determined from our photoionization models of ten PNe.  In most cases, the abundances agree to within 25--40\%.  The few discrepant abundances serve to illustrate the magnitude of errors in the ICFs.  When uncertainties in the ionic and elemental abundances used in the ICFs are taken into account, the Se and Kr ICFs are uncertain by at least 20\% and 40\%, respectively, and the errors may approach a factor of two in some cases.  It should be noted that the ten PNe we have considered are among the brightest and most well-studied PNe in the Northern Hemisphere.  For fainter objects in our NIR survey, the ICFs may be even more uncertain.  Therefore, when a realistic and rigorous error analysis is applied, our derived Se and Kr abundances are unlikely to be determined with an accuracy of better than 50\%.

A more serious factor in the accuracy of our Se and Kr abundance determinations is the uncertain atomic data governing the ionization equilibrium of these elements (PI cross-sections and rate coefficients for RR, DR, and CT).  We have run Monte Carlo simulations to test the importance of these uncertainties in our models of four PNe, allowing the rate coefficients of these atomic processes to vary within a Gaussian distribution of FWHM 0.5~dex around the adopted values, or a factor of three (1-$\sigma$) uncertainty.  In particular, uncertainties in the PI cross-sections are the most significant, and can lead to errors of up to 0.2--0.25~dex ($\sim60$--80\%) in the predicted line intensities of [\ion{Se}{4}], [\ion{Kr}{3}], [\ion{Kr}{4}], and [\ion{Kr}{5}].  Uncertainties in the CT rate coefficients are the next most important, especially for the lower ionization states, and can result in errors of up to 0.15 dex (40\%) in the [\ion{Kr}{3}] intensities.  The rate coefficients for low-$T$~DR are also significant, and can lead to errors of 0.1--0.2 dex (25--60\%) in the predicted intensities of high ionization Se and Kr lines, particularly for low excitation objects.  The uncertainties in the predicted line intensities correspond to uncertainties in the abundances of the parent ions, which propagate into the elemental abundance errors.  Taken together, the atomic data uncertainties alone can result in Se abundance errors of 0.3~dex (a factor of two), and Kr abundance uncertainties of up to 0.2--0.25~dex.

Therefore, calculations and measurements of PI cross-sections and RR, DR, and CT rate coefficients for Se and Kr ions are needed to improve the accuracy to which the abundances of these elements may be determined in PNe.  Furthermore, most other \emph{n}-capture elements also lack such atomic data, and this is a serious impediment to studying \emph{s}-process enrichments in PNe.  Other \emph{n}-capture elements, including Br ($Z=35$), Rb ($Z=37$), and Xe ($Z=54$) have been detected in NGC~7027 (PB94; Zhang et al.\ 2005) and other bright PNe (Sharpee et al.\ 2003; Liu et al.\ 2004a; Zhang et al.\ 2006).  Accurately determining the abundances of these elements, particularly Xe, in conjunction with Se and/or Kr would provide crucial details of the physical conditions under which \emph{s}-process nucleosynthesis operates in PN progenitor stars.

An investigation of the unknown Se and Kr atomic data is difficult and time-consuming, and was not attempted for our current study.  However, one of us (NCS) has begun a laboratory astrophysics study of Se, Kr, and Xe, which will provide much of the unknown atomic data for these elements.  This study will utilize both theoretical calculations and experimental measurements to determine the data.  In fact, the PI cross-sections of Se$^{+}$ and Se$^{++}$ near threshold have already been measured at the Advanced Light Source in Berkeley, CA, and are discussed in Sterling (2006b).

While the atomic data uncertainties must be acknowledged as an important source of error in our Se and Kr abundance determinations at this stage, our model-derived ICFs  are as accurate as is currently possible.  In Paper~II, we will apply these ICFs to our full sample of PNe to derive Se and Kr abundances or upper limits.  The large fraction of PNe in which the Se and/or Kr NIR lines have been detected (81 of 120 objects) enables us to determine \emph{n}-capture element abundances in a large sample of Galactic PNe for the first time; indeed, our survey will increase the number of PNe with known \emph{n}-capture element abundances by more than a factor of ten.  With this data, we provide in Paper~II the first broad characterization of \emph{s}-process enrichments in PNe, which reveals their dredge-up histories and importance to the evolution of trans-iron elements in the Galaxy.

\acknowledgements

We are indebted to several people whose contributions have made this work possible.  We are especially grateful to K.\ Butler for calculating the [\ion{Se}{4}]~2.287~$\mu$m effective collision strength, and G.\ Ferland for helpful discussions and suggestions for expanding the Cloudy atomic database.  We thank K.\ Fournier for providing DR rate coefficients for Kr (and M.\ Mattioli for collecting this data), L.\ Koesterke and T.\ Rauch for computing stellar atmosphere models for IC~418, M.\ Richter and A.\ Sorokin for sending us the Kr$^0$ PI cross-section data, and F.\ Robicheaux for providing Se$^0$ PI cross-section data.  We also acknowledge D.\ W.\ Savin and G.\ Shields for their careful reading of this manuscript and helpful comments.  This work has been supported by NSF grants AST~97-31156 and AST~04-06809.

\appendix

\section{APPENDIX: SE AND KR LINE IDENTIFICATIONS}

In this appendix, we discuss the identifications of the optical and NIR Kr and Se emission lines listed in Table~2.  These lines have been associated with several different species in the literature, if identified at all.  For example, [\ion{Kr}{4}]~$\lambda$5346.0 has previously been identified as \ion{C}{3} (Liu et al.\ 2004a) and \ion{S}{2} (Peimbert et al.\ 2004; Hyung et al.\ 1994a, 1995), while [\ion{Kr}{4}]~$\lambda$5867.7 has been associated with a permitted \ion{Al}{2} triplet (Sharpee et al.\ 2003) as well as \ion{He}{2} Pfund~29.

Considering the confusion in the literature regarding the identity of these lines, we deemed it worthwhile to investigate all possible identifications for the putative Se and Kr lines.  We have determined that, at least for the 10 PNe we have modeled in detail, the evidence we have accumulated strongly indicates that the lines listed in Table~2 are indeed from Se and Kr.

In the following, we consider each line separately.  To illustrate our identification method, we present our investigation of $\lambda$5346.0 in detail.  For the other features, we provide less detail, but discuss important line blends and how the contaminants were removed.

\subsection{The $\lambda$5346.0 Feature}

We have used the Atomic Line List v2.04\footnote{P.\ A.\ M.\ van~Hoof (1999), http://www.pa.uky.edu/$\sim$peter/atomic/ } compiled by P.\ A.\ M.\ van~Hoof to search for alternate identifications of [\ion{Kr}{4}]~$\lambda$5346.0.  We searched a region $\pm0.5$~\AA\ around the central wavelength, and considered all elements through $Z=30$.  We excluded forbidden transitions between levels higher than 2 eV above the ground state, as well as intercombination lines of elements other than C, N, and O.  Furthermore, we did not consider permitted transitions of rare or highly refractory elements (e.g., Cr, Cu, Mg, Mn).  The remaining possible identifications are listed in Table~A1, along with the energies of the upper and lower levels of the transition, and the wavelengths of other lines from the same multiplet.  This list includes the tentative \ion{S}{2} and \ion{C}{3} identifications that have appeared in the literature (Hyung et al.\ 1994a, 1995; Liu et al.\ 2004a; Peimbert et al.\ 2004).

With the above constraints, all of the possible alternative identifications of $\lambda$5346.0 are permitted lines.  If any of these identifications are correct, other lines from the same multiplet should be seen.  Of course, some of these may be undetected if they are relatively weak, and for many of the species in Table~A1, the relative strengths of other multiplet members are unknown.  Nevertheless, it is noteworthy that we do not find \textit{any} of the other multiplet members in the optical spectra of the ten modeled PNe.

In particular, we can dismiss the identification as \ion{C}{3}~$\lambda$5345.9.  According to Wiese et al.\ (1996, hereafter WFD96), in LTE \ion{C}{3}~$\lambda$5359.9 should be nearly twice the strength of $\lambda$5345.9.  Although this region of the spectra is free from other features, this line is undetected in all of the optical spectra we have utilized.  \ion{C}{3}~$\lambda$5337.4, half the strength of $\lambda$5345.9 in LTE (WFD96), is also unseen.  These lines are also undetected in echelle spectra of eight of the modeled PNe (all but NGC~6826 and NGC~6886) obtained by Dinerstein et al.\ (in preparation; see also French et al.\ 2000), despite the presence of $\lambda$5346.0 in each.

Analogously, \ion{N}{1}~$\lambda$5346.5 is the weakest member of its triplet in LTE (WFD96), but the strongest line, $\lambda$5310.4, is undetected in the optical spectra of all 10 PNe.

While the relative strengths of multiplet members for other lines in Table~A1 are not known (to our knowledge), we consider their non-detection to be evidence that these species do not contribute significantly to the flux of $\lambda$5346.0.  The only plausible identification that remains is [\ion{Kr}{4}]~$\lambda$5346.0, which is strengthened by the presence of another [\ion{Kr}{4}] line at 5867.7~\AA.

\subsection{The $\lambda$5867.7 Feature}

As with [\ion{Kr}{4}]~$\lambda$5346.0, we compiled all possible alternate identifications of \linebreak\ [\ion{Kr}{4}]~$\lambda$5867.7 with the Atomic Line List.  We excluded \ion{Al}{2}~$\lambda\lambda$5867.6, 5867.8, 5867.9 (Sharpee et al.\ 2003) based on the non-detection of their multiplet members at $\lambda\lambda$5853.8 and 5861.7 (although the latter may be lost in a blend with [\ion{Mn}{5}]~$\lambda$5861.0 in some objects).  Other possible contaminants were ruled out using the same argument, with one exception.

\ion{He}{2}~$\lambda$5869.0 (Pfund~29) is a potential contaminant to [\ion{Kr}{4}]~5867.7, especially in low resolution spectra (such as that of Liu et al.\ 2004a,b and Zhang et al.\ 2005).  Therefore, we have subtracted the contribution of this line to the reported $\lambda$5867.7 intensities in the optical data references, by using the intensities of nearby isolated \ion{He}{2} Pfund series members and the Fortran program of Storey \& Hummer (1995).  Note that we perform this subtraction even in spectra with sufficiently high resolution to resolve these features if the \ion{He}{2} intensity was not given separately from that of [\ion{Kr}{4}].  In these calculations, we used the [\ion{O}{3}] temperatures and averaged electron densities derived from the optical spectra.  The remaining flux, which we attribute to [\ion{Kr}{4}], is given in Table~2.  This correction reduces $I$($\lambda$5867.7) by as little as 8\% (NGC~6790) to up to 40\% (NGC~6886), and hence [\ion{Kr}{4}] is the primary contributor to this feature in all the PNe we have modeled.

\subsection{The $\lambda$6826.7 Feature}

Baluteau et al.\ (1995) listed two potential contaminants to [\ion{Kr}{3}]~$\lambda$6826.7: \ion{He}{1}~$\lambda$6827.9 and \ion{C}{1}~$\lambda$6828.1.  These transitions are the most likely alternate identifications for this line, and other possibilities have been excluded as in \S A.1.

\ion{C}{1}~$\lambda$6828.1 is a singlet; however, we deem this an unlikely contaminant since other lines from the same upper level ($\lambda\lambda$6294.1, 6656.5, 7366.8) are not seen in the optical spectra of the 10 PNe we have modeled.

On the other hand, \ion{He}{1}~$\lambda$6827.9 is likely to contribute to this line in low resolution spectra, and other lines from the 3$^1S$--$n^3P$ series (P\'{e}quignot \& Baluteau 1988; Smits 1991) have been detected in the modeled PNe.  Using the intensities of neighboring lines in this series and the Smits (1991) tables of theoretical \ion{He}{1} line intensities at $T_{\rm e}=10^4$~K and $n_{\rm e}=10^4$~cm$^{-3}$, we have removed the contribution of \ion{He}{1}~$\lambda$6827.9 to the $\lambda$6826.7 feature.  In some cases (NGC~6790, NGC~6826, NGC~6884, NGC~7662) this correction can be significant ($\sim$40\%), while for the other PNe of our sample, it reduces $I$($\lambda$6826.7) by less than 20\%.  We attribute the remaining flux (listed in Table~2) to [\ion{Kr}{3}]~$\lambda$6826.7.

In addition to nebular contaminants, [\ion{Kr}{3}]~$\lambda$6826.7 can also be affected by telluric OH emission at $\lambda$6827.5 (Hanuschik 2003; Osterbrock et al.\ 1996).  For this reason, the intensity of [\ion{Kr}{3}]~$\lambda$6826.7 from low resolution spectra (e.g., Liu et al.\ 2004a,b; Zhang et al.\ 2005) is likely to be more uncertain than that of 2.199~$\mu$m.  This telluric feature may also affect the $\lambda$6826.7 flux in two other PNe exhibiting [\ion{Kr}{4}] emission, NGC~2022 and NGC~6818 (Tsamis et al.\ 2003).  These spectra have resolutions of 4.5~\AA\ FWHM at this wavelength, which is sufficiently low that blending with telluric features may be important.  When we attempted to model the Kr features in these two PNe, we found that the [\ion{Kr}{3}]~$\lambda$6826.7 intensity was far too strong (by more than an order of magnitude) to be consistently fit with the [\ion{Kr}{4}] lines.  Since no \ion{He}{1} 3$^1S$--$n^3P$ series lines were detected in these two PNe, the discrepancy may be due to incomplete removal of the telluric emission near $\lambda$6826.7, or an instrumental artifact.  We did not attempt to model these two PNe further.

\subsection{The 2.199~$\mu$m Feature}

Identified as [\ion{Kr}{3}] by Dinerstein (2001), this line had long been a mystery to infrared astronomers.  It was first detected in a PN by Treffers et al.\ (1976), and Geballe et al.\ (1991) unsuccessfully attempted to identify it (and the 2.287~$\mu$m feature) with transitions of elements from H to Ca and Fe.  While H$_2$~3-2~$S$(3)~2.201~$\mu$m possibly contributes to the flux of this line in some PNe, the detection of the 2.199~$\mu$m feature in PNe that do not exhibit molecular hydrogen indicates that it has an atomic nature, as noted by Geballe et al.

Cox et al.\ (2002) identified this line as \ion{N}{1}~2.199~$\mu$m in NGC~7027, based on the detection of other \ion{N}{1} lines at 2.194 and 2.148~$\mu$m.  However, we do not detect the latter \ion{N}{1} lines in \textit{any} of the 103 PNe we observed, nor are they evident in any of the literature spectra we have utilized, including those of NGC~7027 (Geballe et al.\ 1991; Hora et al.\ 1999; Lumsden et al.\ 2001).  Examining Figure~1 of Cox et al.\ (2002), one sees that \ion{N}{1}~2.194~$\mu$m is much weaker than the 2.199~$\mu$m line, and \ion{N}{1}~2.148~$\mu$m lies in a region of their spectrum which is extremely noisy, and hence its flux is quite uncertain.  This suggests that \ion{N}{1} is a minor contaminant to [\ion{Kr}{3}]~2.199~$\mu$m even in NGC~7027; the absence of NIR \ion{N}{1} features in all of the other PNe in our sample indicates that the \ion{N}{1} contribution to this line is negligible in most cases, and we do not consider it further.

Similarly, we find that other transitions of H through Zn within 0.005~$\mu$m of the 2.199~$\mu$m line are unlikely contaminants, based on the non-detections of permitted line multiplet members, and the lack of forbidden transitions between low-lying energy levels in this wavelength region.  Therefore, the only important contaminant to [\ion{Kr}{3}]~2.199~$\mu$m is H$_2$~3-2~$S$(3)~2.201~$\mu$m.

Of the ten objects considered in this paper, four exhibit NIR H$_2$ emission (IC~5117, NGC~6741, NGC~6886, and NGC~7027).  The contribution of H$_2$ to the 2.199~$\mu$m flux depends on the H$_2$ excitation mechanism.  For example, models of Black \& van~Dishoeck (1987) show that the strengths of H$_2$~3-2 lines in the $K$~band are negligible if the H$_2$ is thermally excited, but may be a significant fraction (30--40\%) of the H$_2$~1-0~$S$(0)~2.224~$\mu$m flux under fluorescent excitation.  In principle, the H$_2$ excitation mechanism can be ascertained from the flux ratio of H$_2$~2-1~$S$(1)~2.248~$\mu$m relative to H$_2$~1-0~$S$(0)~2.224~$\mu$m, which we denote as $F$(2.248/2.224).  The canonical fluorescent-excitation model, Model~14 of Black \& van~Dishoeck, predicts that $F$(2.248/2.224)~=~1.22, while their thermal excitation model (S2) predicts $F$(2.248/2.224)~=~0.38.  The four PNe considered in this paper that exhibit H$_2$ emission all have $F$(2.248/2.224)~$<0.65$, indicating that the observed H$_2$ is collisionally excited, and hence contamination of the [\ion{Kr}{3}] flux (as well as [\ion{Se}{4}]) by H$_2$ is not expected to be important in these objects.

\subsection{The 2.287~$\mu$m Feature}

First identified as [\ion{Se}{4}] by Dinerstein (2001), this line also may be contaminated by molecular hydrogen (H$_2$~3-2~$S$(2)~2.287~$\mu$m), but in \S A.4 we show this to be unimportant in the ten objects we consider in this paper.  Our search for additional contaminants of this line revealed no other likely identifications, in agreement with the conclusions of Geballe et al.\ (1991).  We searched for transitions from H through Zn within 0.005~$\mu$m of 2.287$~\mu$m, and the only likely candidates were permitted lines for which other members of the multiplet were in all cases undetected.

\subsection{The $\lambda\lambda$6256.1, 8243.4 Features}

These two wavelengths correspond to the strongest optical [\ion{Kr}{5}] lines.  Since $\lambda$6256.1 was detected in only one PN in our sample (NGC~7027) and $\lambda$8243.4 not at all, we have not attempted detailed investigations of the identities of these features.  We only note that one must use caution in identifying $\lambda$6256.1 with [\ion{Kr}{5}] due to the nearby lines \ion{C}{2}~$\lambda\lambda$6256.2, 6256.5, and that [\ion{Kr}{5}]~8243.4 is prone to blending with telluric emission, as well as \ion{N}{1}~$\lambda$8242.4 and \ion{O}{3}~$\lambda$8244.1 (Baluteau et al.\ 1995).

\section{APPENDIX: SE AND KR ATOMIC DATA}

In this appendix, we describe the adopted Se and Kr atomic data used in our Cloudy and XSTAR models.  The atomic data governing the ionization equilibrium of these elements, and indeed all \emph{n}-capture elements, are poorly known.  In most cases, the photoionization (PI) cross-sections and rate coefficients for radiative recombination (RR), dielectronic recombination (DR), and charge transfer (CT) have not been calculated or experimentally measured for Se and Kr ions.  For this reason, we have been forced to use approximate atomic data for these processes in order to model the ionization balance of Se and Kr.  In the case of Kr, multiple ionization stages have been observed in the 10 PNe listed in Table~1, which allows us to adjust the uncertain Kr atomic data so that the line intensities are satisfactorily reproduced by our models.  However, no other Se ions have been detected in PNe, and we are unable to empirically correct the Se atomic data.

\subsection{Energy Levels, Transition Probabilities, and Effective Collision Strengths}

Unlike the atomic data controlling the ionization balance of Se and Kr, emission line data has been determined for the observed ions of these elements.  Transition probabilities $A_{\rm ij}$ and effective collision strengths $\Omega_{\rm ij}$ have been determined for \ion{Se}{4}, \ion{Kr}{3}, \ion{Kr}{4}, and \ion{Kr}{5}.  In Table~B1 we list the references from which these data are taken.

All energy level data are taken from the NIST Atomic Spectra Database\footnote{National Institute of Standards and Technology Atomic Spectra Database v3.0; see http://physics.nist.gov/PhysRefData/ASD/index.html}, as are the ionization potentials (IPs) with the exception of Kr$^{10+}$--Kr$^{16+}$, whose IPs are taken from the calculations of Bi\'{e}mont et al.\ (1999).

\subsection{Photoionization Cross-Sections and Radiative Recombination Rate Coefficients}  

PI cross-sections are unknown for all Se and Kr ions except Kr$^0$ (Richter et al.\ 2003) and Se$^0$ (Chen \& Robicheaux 1994, who calculated the cross-section for only a small energy range near threshold).

The classical Kramers (1923) formula,
\begin{equation}
\sigma_{\rm th} = 2^63^{-1.5}\alpha \pi a_0^2(E_{\rm Ryd}/E_{\rm th})(N_{\rm e}/n),
\end{equation}
can be used to estimate the PI cross-section at the threshold energy (i.e., ionization potential).  In this equation, $\sigma_{\rm th}$ is the threshold cross-section, $\alpha$ the fine-structure constant, $a_0$ the Bohr radius, $E_{\rm Ryd}$ the Rydberg energy, $E_{\rm th}$ the threshold energy, $N_{\rm e}$ the number of valence electrons, and $n$ the principal quantum number of the valence electrons.  Gould (1978) has shown that this equation provides a reasonably good approximation (to within a factor of two) of the threshold PI cross-sections of first and second row elements.  However, Equation~(B1) does not provide information regarding the functional dependence of the cross-sections on photon energy.  Therefore, we simply assumed that the PI cross-sections have power law forms above the threshold energies (Osterbrock 1989, pp.\ 34--36):
\begin{equation}
\sigma_{\rm pi}(E) = \sigma_{\rm th}(E/E_{\rm th})^{-2}.
\end{equation}

Although $\sigma_{\rm pi}(E)$ is known for Kr$^0$ and (partially) for Se$^0$, for consistency we have used cross-sections from Equations~(B1) and (B2) for these species in Cloudy and XSTAR.  With XSTAR, we have tested the sensitivity of our results to the adopted form of the Kr$^0$ and Se$^0$ PI cross-sections.  We smoothed the Se$^0$ and Kr$^0$ cross-sections (Chen \& Robicheaux 1994; Richter et al.\ 2003) using the resonance-averaged photoionization (RAP) cross-section method of Bautista et al.\ (1998).  For energies not considered by Chen \& Robicheaux, we assumed $\sigma_{\rm pi} (E)$ is a power-law with index $-2$, as in Equation~(B2).  The Kr$^0$ and Se$^0$ PI cross-sections were entered into XSTAR as energy-$\sigma_{\rm pi}$ pairs which are linearly interpolated to reproduce the cross-sections.  As expected, we find that the predicted Se and Kr line intensities depend negligibly ($<2$\%) on the adopted  Se$^0$ and Kr$^0$ PI cross-sections.  The reason is that neutral Se and Kr are trace species in our models, and hence their fractional abundances do not significantly affect the derived elemental abundances.

We have determined $\sigma_{\rm pi}(E)$ for all terms in the ground principal quantum number shell of each Se and Kr ion with Equations (B1) and (B2), and used the Milne relation (e.g., Osterbrock \& Ferland 2006, p.\ 401) to determine the RR rate coefficient $A_{\rm rr}$ into this shell.  For higher principal quantum number shells, we adopted hydrogenic recombination rate coefficients (Seaton 1959).  Adding the $A_{\rm rr}$ for the ground and higher principal quantum number shells of an ion gives the total $A_{\rm rr}$.  The $A_{\rm rr}$ of high ions of Se and Kr (greater than 5--6 times ionized) are underestimated, due to incomplete energy level data for these species (the lack of energy level data in the ground principal quantum number shell prevents the calculation of RR rate coefficients into these states).  However, since these ionization states are usually sparsely populated in photoionized nebulae, this should have a negligible effect on our results.

The dependence of the RR rate coefficient on temperature is often approximated as a power law (e.g., Aldrovandi \& P\'{e}quignot 1973), 
\begin{equation}
\alpha_{\rm r}(T) = A_{\rm rr} t^{- \eta },
\end{equation}
where $t$ is the electron temperature in units of 10$^4$~K.  We adopt $\eta=0.7$, which has been found to be a good approximation even for heavy element ions (e.g., Mazzitelli \& Mattioli 2002).

The Se and Kr PI cross-sections and RR rate coefficients we use in Cloudy and XSTAR are given by Equations (B1)--(B3), with the fit parameters listed in Table~B2.

\subsection{Charge Transfer Rate Coefficients}

CT rate coefficients are not known for any ions of Se and Kr.  The heaviest elements with reliable CT data are Ni and Cu (Kingdon \& Ferland 1996)\footnote{While these authors derived CT data for Zn, the rate coefficients of Zn$^{3+}$ and Zn$^{4+}$ are likely to be underestimates due to the incomplete energy level data of these ions (since electrons are expected to be captured into energy levels higher than the limit of the published levels).}.  Therefore, we assume that the rate coefficients of the first four ions of Se and Kr are the same as the averaged rate coefficients of similar charge states of Ni and Cu.

For higher Se and Kr ions, we adopt the formula of Ferland et al.\ (1997):
\begin{equation}
\alpha_{\rm ct} = 1.92\times10^{-9}\zeta ~~(\mathrm{cm}^3~\mathrm{s}^{-1}),
\end{equation}
where $\alpha_{\rm ct}$ is the rate coefficient, and $\zeta$ is the ion charge.

\subsection{Dielectronic Recombination Rate Coefficients}

DR is the dominant recombination process for many species in photoionized nebulae.  There are essentially two different types of DR which can be important in ionized plasmas: high-temperature DR, where core relaxation stabilizes the ion, and low-temperature DR, where the recombining electron is captured into a low-lying resonance state (Dopita \& Sutherland 2003, p.\ 108).

In photoionized nebulae, low-$T$ DR dominates high-$T$ DR in almost all cases.  However, low-$T$ DR rate coefficients are unknown for most third row and higher elements, because the energies of the low-lying autoionizing resonance states have not been experimentally measured (Ferland et al.\ 1998; Ferland 2003).  The unknown energies of these states precludes theoretical calculations of the low-$T$ DR rate coefficients for heavy elements.  In principle, the rate coefficients can be determined experimentally with electron-ion beam merging techniques such as those of Savin et al.\ (1999, 2006), but it is difficult to use these methods to measure DR rate coefficients for the low charge-to-mass species found in photoionized nebulae, and to date no such measurements have been performed.

In Cloudy, low-$T$ DR is approximated for all third row and higher elements in a manner depending on the ionization stage.  For the first four ions, the rate coefficients are assumed to be equal to the averaged rate coefficients of the same charge states of C, N, and O.  For higher ions, the rate coefficient is given by
\begin{equation}
\alpha_{\rm dr}^{\mathrm{Low-T}} = 3.0\times10^{0.1\zeta -12} ~~(\mathrm{cm}^3~\mathrm{s}^{-1}),
\end{equation}
where $\zeta$ is the ion charge.  We have adopted these approximations for Se and Kr in Cloudy, and added them to the atomic database of XSTAR for all third row and heavier elements.

Although high-$T$ DR is less important in photoionized nebulae, rate coefficients for this process have been determined for all Kr ions by Fournier et al.\ (2000).  We fit these data with the formula of Aldrovandi \& P\'{e}quignot (1973, their equation 5):
\begin{equation}
\alpha_{\rm dr}^{\mathrm{High-T}} = A_{\rm di}T_{\rm e}^{-1.5}e^{-T_0/T_{\rm e}}(1+B_{\rm di}e^{-T_1/T_{\rm e}}),
\end{equation}
where $T_{\rm e}$ is the electron temperature (in K), and $A_{\rm di}$, $T_0$, $B_{\rm di}$, and $T_1$ are the fitting parameters (listed in Table~B3).  The high-$T$ DR rate coefficients for Se are unknown, and hence we assumed the same rate coefficients as Kr ions of the same charge.

\subsection{Collisional Ionization Cross-Sections}

Like high-$T$ DR, collisional ionization (CI) plays a minor role in most photoionized nebulae, since the plasma is at too low a temperature ($\sim$1~eV) to ionize atoms.  For completeness, we have added the Kr CI data of Loch et al.\ (2002)\footnote{see http://www-cfadc.phy.ornl.gov/data\_and\_codes/aurost/aurost\_ioniz/kr-iso-nuclear.html} to Cloudy and XSTAR.  We fit the data with the formula of Voronov (1997):
\begin{equation}
<\sigma_{\rm ci}v> = A_{\rm ci}\frac{(1+PU^{0.5})}{(X+U)}U^Ke^{-U},
\end{equation}
where $<\sigma_{\rm ci}v>$ is the rate coefficient, $U=E_{\rm th}/T_{\rm e}$, and $A_{\rm ci}$, $P$, $X$, and $K$ are the fit parameters (given in Table~B3).  The CI rate coefficients are not known for Se, and hence we assume Se ions have the same rate coefficients as Kr ions of the same charge.

\clearpage

\begin{deluxetable}{lccccc}
\tablecolumns{6}
\tabletypesize{\footnotesize}
\tablewidth{0pc} 
\tablecaption{Optical and Near-Infrared Data References}
\tablehead{
\colhead{Object} & \colhead{Optical} & \colhead{Spectral} & \colhead{} & \colhead{NIR Data} & \colhead{Other} \\
\colhead{Name} & \colhead{Data Ref.} & \colhead{Res.\ (\AA)\tablenotemark{a}} & \colhead{$c$(H$\beta$)} & \colhead{Ref.} & \colhead{Ref.\tablenotemark{b}}}
\startdata
IC418 & Sharpee et al.\ 2003 & 0.1--0.3 & 0.34 & Paper II & Pottasch et al.\ 2004 \\
IC 5117 & Hyung et al.\ 2001 & 0.05--0.15 & 1.40 & Rudy et al.\ 2001 & \nodata \\
NGC6572 & Liu et al.\ 2004a,b & 1.5--3.5 & 0.48 & Paper II & Hyung et al.\ 1994a\tablenotemark{c} \\
NGC6741 & Liu et al.\ 2004a,b & 1.5--3.5 & 1.15 & Paper II & \nodata \\
NGC6790 & Liu et al.\ 2004a,b & 1.5--3.5 & 1.10 & Paper II & Aller et al.\ 1996\tablenotemark{c} \\
NGC6826 & Liu et al.\ 2004a,b & 1.5--3.5 & 0.06 & Paper II & \nodata \\
NGC6884 & Liu et al.\ 2004a,b & 1.5--3.5 & 1.00 & Paper II & \nodata \\
NGC6886 & Hyung et al.\ 1995 & 0.05--0.15 & 0.90 & Paper II & Pottasch \& Surendiranath 2005 \\
NGC 7027 & Zhang et al.\ 2005 & 1.5--4.5 & 1.37 & Geballe et al.\ 1991 & \nodata \\
NGC 7662 & Liu et al.\ 2004a,b & 1.5--3.5 & 0.18 & Geballe et al.\ 1991 & \nodata \\
\tableline
\enddata
\tablecomments{Optical and NIR data references are given, along with the optical spectral resolutions and extinction coefficients $c$(H$\beta$).}
\tablenotetext{a}{The range of FWHM resolution of the optical spectrum.}
\tablenotetext{b}{Reference for abundances, UV, and IR data if not from optical data reference.}
\tablenotetext{c}{[\ion{Kr}{4}]~$\lambda$5867.7 is taken from this reference and scaled to [\ion{Kr}{4}]~$\lambda$5346.0 of Liu et al.\ (2004a,b).  All other abundance, optical, and IR data (except the $K$ band) are from Liu et al.}
\end{deluxetable}
 
\clearpage

\begin{deluxetable}{lcccccc}
\tablecolumns{7}
\tablewidth{0pc} 
\tablecaption{Dereddened Se and Kr Line Intensities Relative to H$\beta$ ($I$(H$\beta)=10^4$)}
\tablehead{
\colhead{PN} & \colhead{$[$\ion{Se}{4}$]$} & \colhead{$[$\ion{Kr}{3}$]$} & \colhead{$[$\ion{Kr}{3}$]$} & \colhead{$[$\ion{Kr}{4}$]$} & \colhead{$[$\ion{Kr}{4}$]$} & \colhead{$[$\ion{Kr}{5}$]$} \\
\colhead{Name} & \colhead{2.287~$\mu$m} & \colhead{$\lambda$6826.7\tablenotemark{a}} & \colhead{2.199~$\mu$m} & \colhead{$\lambda$5346.0} & \colhead{$\lambda$5867.7\tablenotemark{b}} & \colhead{$\lambda$6256.1}}
\startdata
IC 418 & \nodata & 2.90$\pm$0.87\tablenotemark{c} & 6.99$\pm$0.54 & 0.35$\pm$0.11\tablenotemark{c} & 0.35$\pm$0.11\tablenotemark{c} & \nodata \\
IC 5117 & 31.5$\pm$3.4 & \nodata & 4.23$\pm$1.06 & 12.9$\pm$3.9 & 14.5$\pm$4.4 & \nodata \\
NGC 6572 & 8.87$\pm$0.51 & 1.80$\pm$0.54 & 2.49$\pm$0.61 & 4.70$\pm$0.94 & 6.40$\pm$1.92 & \nodata \\
NGC 6741 & 9.00$\pm$2.23 & 3.70$\pm$0.37 & 5.40$\pm$1.38 & 6.90$\pm$1.38 & 5.50$\pm$1.10 & \nodata \\
NGC 6790 & 8.61$\pm$0.91 & 0.45$\pm$0.14 & \nodata & 5.30$\pm$1.06 & 5.20$\pm$1.04 & \nodata \\
NGC 6826 & 5.41$\pm$1.55 & 0.93$\pm$0.28 & \nodata & 2.30$\pm$0.69 & 4.50$\pm$1.35 & \nodata \\
NGC 6884 & 20.3$\pm$3.0 & 0.83$\pm$0.25 & \nodata & 9.50$\pm$0.95 & 11.4$\pm$1.1 & \nodata \\
NGC 6886 & 14.6$\pm$1.6 & \nodata & 5.89$\pm$1.19 & 8.00$\pm$2.4\tablenotemark{c} & 5.40$\pm$1.6\tablenotemark{c} & \nodata \\
\enddata
\tablecomments{Error bars are 1-$\sigma$ uncertainties, and account only for uncertainties in the flux measurements.  Line intensities are taken from the references in Table~1.}
\tablenotetext{a}{Corrected for \ion{He}{1}~$\lambda$6827.9 contamination as described in Appendix A.3.}
\tablenotetext{b}{Corrected for \ion{He}{2}~$\lambda$5869.0 contamination as described in Appendix A.2.}
\tablenotetext{c}{An uncertainty of 30\% in the line intensity is assumed, since errors were not reported in the optical data reference.}
\end{deluxetable}

\clearpage

\begin{deluxetable}{lccccccc}
\tablecolumns{8}
\tablewidth{0pc} 
\tabletypesize{\footnotesize}
\tablecaption{Cloudy Input Parameters}
\tablehead{
\colhead{Parameter} & \colhead{IC 418\tablenotemark{a}} & \colhead{IC 418\tablenotemark{b}} & \colhead{IC 5117} & \colhead{NGC 6572} & \colhead{NGC 6741} & \colhead{NGC 6790} & \colhead{Solar\tablenotemark{c}}}
\startdata
$T_{\rm eff}$ (10$^3$~K) & 38.8$\pm$3.0 & 37.3$\pm$2.0 & 125.25$\pm$10.0 & 78.71$\pm$7.0 & 152.49$^{+10.0}_{-20.0}$ & 96.65$\pm$10.0 & \nodata \\
log $g$ & 3.4 & 3.4 & 8.0 & 5.5 & 5.84 & 5.5 & \nodata \\
$L/L_{\odot}$ & 4390 & 7175 & 3300 & 3820 & 910 & 5800 & \nodata \\
$n_{\rm H}$ (10$^3$ cm$^{-3}$) & 13.0 & 13.0 & 57.4 & 17.33 & 5.5 & 50.75 & \nodata \\
$R_{\rm in}$ (log pc) & $-3.0$ & $-3.0$ & $-2.1$ & $-2.0$ & $-2.24$ & $-2.3$ & \nodata \\
$R_{\rm out}$ (log pc) & $-1.34$ & $-1.45$ & $-1.61$ & $-1.44$ & $-1.26$ & $-1.74$ & \nodata \\
$D$ (kpc)\tablenotemark{d} & 1.6 & 2.1 & 14.0 & 2.6 & 6.5 & 11.2 & \nodata \\
He/H & 0.106 & 0.141 & 0.100 & 0.104 & 0.105 & 0.094 & 0.085 \\
\enddata
\tablecomments{Input parameters for final Cloudy models.  The stellar effective temperature, gravity, and luminosity (relative to solar), and nebular hydrogen density, inner and outer radii, and implied distance are given in the first seven rows.  Abundances follow, with He/H on a linear scale, and the other elements given as $12+$~log(X/H).}
\tablenotetext{a}{IC 418 model with Rauch NLTE model atmosphere and no mass loss.}
\tablenotetext{b}{IC 418 model with stellar atmosphere including mass-loss (see text, \S3.1.1).}
\tablenotetext{c}{From Asplund et al.\ (2005)}
\tablenotetext{d}{Distance implied from the model, calculated by comparing the predicted H$\beta$ luminosity to the observed H$\beta$ flux from Cahn et al.\ (1992).  Since we did not attempt to optimize $L$(H$\beta$) in our models, these values should not be viewed as accurate estimates of the distances to the modeled PNe.}
\end{deluxetable}

\clearpage

\begin{deluxetable}{lcccccc}
\tablecolumns{7}
\tablewidth{0pc} 
\tablecaption{Cloudy Input Parameters}
\tablehead{
\colhead{Parameter} & \colhead{NGC 6826} & \colhead{NGC 6884} & \colhead{NGC 6886} & \colhead{NGC 7027} & \colhead{NGC 7662} & \colhead{Solar\tablenotemark{a}}}
\startdata
$T_{\rm eff}$ (10$^3$~K) & 55.6$\pm$5.0 & 114.82$\pm$10.0 & 164.4$\pm$20.0 & 173.3$^{+15.0}_{-30.0}$ & 125.6$\pm$10.0 & \nodata \\
log $g$ & 5.0 & 5.7 & 7.0 & 6.0 & 5.7 & \nodata \\
$L/L_{\odot}$ & 10600 & 1520 & 690 & 4900 & 2470 & \nodata \\
$n_{\rm H}$ (10$^3$ cm$^{-3}$) & 1.6 & 7.0 & 8.1 & 55.0 & 3.21 & \nodata \\
$R_{\rm in}$ (log pc) & $-1.6$ & $-2.7$ & $-2.79$ & $-2.05$ & $-1.60$ & \nodata \\
$R_{\rm out}$ (log pc) & $-0.85$ & $-1.46$ & $-1.58$ & $-1.67$ & $-1.43$ & \nodata \\
$D$ (kpc)\tablenotemark{b} & 4.1 & 6.1 & 5.4 & 3.7 & 1.7 & \nodata \\
He/H & 0.099 & 0.099 & 0.107 & 0.100 & 0.103 & 0.085 \\
\enddata
\tablecomments{Input parameters for final Cloudy models.  The stellar effective temperature, gravity, and luminosity (relative to solar), and nebular hydrogen density, inner and outer radii, and implied distance are given in the first seven rows.  Abundances follow, with He/H on a linear scale, and the other elements given as $12+$~log(X/H).}
\tablenotetext{a}{From Asplund et al.\ (2005)}
\tablenotetext{b}{Distance implied from the model, calculated by comparing the predicted H$\beta$ luminosity to the observed H$\beta$ flux from Cahn et al.\ (1992).  Since we did not attempt to optimize $L$(H$\beta$) in our models, these values should not be viewed as accurate estimates of the distances to the modeled PNe.}
\end{deluxetable}

\clearpage

\begin{deluxetable}{lc|cc|cc|cc|cc}
\tablecolumns{10}
\tabletypesize{\scriptsize}
\tablewidth{0pc} 
\tablecaption{Cloudy Predicted vs. Observed Diagnostics and Line Intensities}
\tablehead{
\multicolumn{2}{c|}{} & \multicolumn{2}{|c|}{IC 418\tablenotemark{a}} & \multicolumn{2}{|c|}{IC 418\tablenotemark{b}} & \multicolumn{2}{|c|}{IC 5117} & \multicolumn{2}{|c}{NGC 6572} \\ \cline{3-4} \cline{5-6} \cline{7-8} \cline{9-10}
\multicolumn{1}{c}{Ion} & \multicolumn{1}{c}{$\lambda$} & \multicolumn{1}{|c}{Pred.} & \multicolumn{1}{c|}{Obs.} & \multicolumn{1}{|c}{Pred.} & \multicolumn{1}{c|}{Obs.} & \multicolumn{1}{|c}{Pred.} & \multicolumn{1}{c|}{Obs.} & \multicolumn{1}{|c}{Pred.} & \multicolumn{1}{c}{Obs.}}
\startdata
$[$\ion{Ar}{3}$]$na & \nodata & 2.23E+02 & 2.71E+02 & 2.19E+02 & 2.71E+02 & 7.66E+01 & 1.35E+02 & 1.03E+02 & 1.69E+02 \\
$[$\ion{Ar}{4}$]$nn & \nodata & 2.00E+00 & 1.20E+00 & 2.03E+00 & 1.20E+00 & 4.10E+00 & 4.18E+00 & 2.23E+00 & 1.98E+00 \\
$[$\ion{Cl}{3}$]$nn & \nodata & 5.07E$-01$ & 5.10E$-01$ & 4.97E$-01$ & 5.10E$-01$ & 2.96E$-01$ & 3.28E$-01$ & 4.54E$-01$ & 4.40E$-01$ \\
$[$\ion{N}{2}$]$na & \nodata & 6.58E+01 & 7.84E+01 & 6.75E+01 & 7.84E+01 & 1.78E+01 & 2.53E+01 & 3.76E+01 & 3.78E+01 \\
$[$\ion{O}{2}$]$na & \nodata & 4.99E+00 & 6.98E+00 & 4.98E+00 & 6.98E+00 & 8.71E$-01$ & 6.95E$-01$ & 2.79E+00 & 3.06E+00 \\
$[$\ion{O}{2}$]$nn & \nodata & 3.26E+00 & 2.37E+00 & 3.28E+00 & 2.37E+00 & 3.64E+00 & 3.10E+00 & 3.32E+00 & 2.48E+00 \\
$[$\ion{O}{3}$]$na & \nodata & 3.12E+02 & 3.08E+02 & 2.98E+02 & 3.08E+02 & 6.83E+01 & 8.83E+01 & 2.94E+01\tablenotemark{c} & 4.30E+01\tablenotemark{c} \\
$[$\ion{S}{2}$]$nn & \nodata & 4.84E$-01$ & 4.70E$-01$ & 4.84E$-01$ & 4.70E$-01$ & 4.46E$-01$ & 4.36E$-01$ & 4.82E$-01$ & 4.70E$-01$ \\
\enddata
\tablecomments{Comparison of predicted and observed line intensities relative to H$\beta$ (on the scale $I$(H$\beta)=100$) and diagnostic ratios used in the spectral fits.  Wavelengths are in air for $\lambda>3000$~\AA\ and vacuum otherwise.  Listed wavelengths are in \AA\ except where noted.  The diagnostic ratios are defined as follows: [\ion{Ar}{3}]na = [\ion{Ar}{3}] ($7135.8+7751.1$)/5191.8 ; [\ion{Ar}{4}]nn = [\ion{Ar}{4}] 4740.2/4711.4; [\ion{Cl}{3}]nn = [\ion{Cl}{3}] 5517.7/5537.9; [\ion{N}{2}]na = [\ion{N}{2}] ($6548.0+6583.5$)/5754.6; [\ion{O}{2}]na = [\ion{O}{2}] ($3726.0+3728.8$)/($7319.5+7330.2$); [\ion{O}{2}]nn = [\ion{O}{2}] 3726.0/3728.8; [\ion{O}{3}]na = [\ion{O}{3}] ($4958.9+5006.8$)/4363.2; and [\ion{S}{2}]nn = [\ion{S}{2}] 6716.4/6730.8.}
\tablenotetext{a}{IC 418 model with Rauch NLTE model atmosphere and no mass loss.}
\tablenotetext{b}{IC 418 model with stellar atmosphere including mass-loss (see text, \S3.1.1).}
\tablenotetext{c}{[\ion{O}{3}]na = [\ion{O}{3}] 4958.9/4363.2.}
\end{deluxetable}
 
\clearpage

\begin{deluxetable}{lc|cc|cc|cc|cc}
\tablecolumns{10}
\tabletypesize{\scriptsize}
\tablewidth{0pc} 
\tablecaption{Cloudy Predicted vs. Observed Diagnostics and Line Intensities}
\tablehead{
\colhead{} & \colhead{} & \multicolumn{2}{|c|}{NGC 6741} & \multicolumn{2}{|c|}{NGC 6790} & \multicolumn{2}{|c|}{NGC 6826} & \multicolumn{2}{|c}{NGC 6884} \\ \cline{3-4} \cline{5-6} \cline{7-8} \cline{9-10}
\multicolumn{1}{c}{Ion} & \multicolumn{1}{c}{$\lambda$} & \multicolumn{1}{|c}{Pred.} & \multicolumn{1}{c|}{Obs.} & \multicolumn{1}{|c}{Pred.} & \multicolumn{1}{c|}{Obs.} & \multicolumn{1}{|c}{Pred.} & \multicolumn{1}{c|}{Obs.} & \multicolumn{1}{|c}{Pred.} & \multicolumn{1}{c}{Obs.}}
\startdata
$[$\ion{Ar}{3}$]$na & \nodata & 9.64E+01 & 1.32E+02 & 7.52E+01 & 9.16E+01 & 2.11E+02 & 2.56E+02 & 1.11E+02 & 1.51E+02 \\
$[$\ion{Ar}{4}$]$nn & \nodata & 1.28E+00 & 1.26E+00 & 3.85E+00 & 4.06E+00 & 9.04E$-01$ & 9.32E$-01$ & 1.42E+00 & 1.56E+00 \\
$[$\ion{Cl}{3}$]$nn & \nodata & 7.47E$-01$ & 7.30E$-01$ & 3.07E$-01$ & 4.07E$-01$ & 1.08E+00 & 1.10E+00 & 6.70E$-01$ & 7.04E$-01$ \\
$[$\ion{N}{2}$]$na & \nodata & 5.66E+01 & 6.48E+01 & 1.90E+01 & 2.27E+01 & 9.01E+01 & 7.91E+01 & 4.75E+01 & 5.25E+01 \\
$[$\ion{O}{2}$]$na & \nodata & 7.58E+00 & 9.47E+00 & 9.64E$-01$ & 1.51E+00 & 2.32E+01 & 1.48E+01 & 5.83E+00 & 7.02E+00 \\
$[$\ion{O}{2}$]$nn & \nodata & 2.63E+00 & 2.04E+00 & 3.62E+00 & 2.45E+00 & 1.85E+00 & 1.48E+00 & 2.89E+00 & 2.17E+00 \\
$[$\ion{O}{3}$]$na & \nodata & 2.39E+01\tablenotemark{a} & 2.67E+01\tablenotemark{a} & 1.77E+01 & 2.24E+01 & 7.14E+01\tablenotemark{a} & 6.71E+01\tablenotemark{a} & 3.20E+01\tablenotemark{a} & 3.78E+01\tablenotemark{a} \\
$[$\ion{S}{2}$]$nn & \nodata & 5.81E$-01$ & 5.70E$-01$ & 4.48E$-01$ & 4.57E$-01$ & 7.21E$-01$ & 7.30E$-01$ & 5.21E$-01$ & 5.32E$-01$ \\
\enddata
\tablecomments{Comparison of predicted and observed line intensities relative to H$\beta$ (on the scale $I$(H$\beta)=100$) and diagnostic ratios used in the spectral fits.  Wavelengths are in air for $\lambda>3000$~\AA\ and vacuum otherwise.  Listed wavelengths are in \AA\ except where noted.  The diagnostic ratios are defined as follows: [\ion{Ar}{3}]na = [\ion{Ar}{3}] ($7135.8+7751.1$)/5191.8 ; [\ion{Ar}{4}]nn = [\ion{Ar}{4}] 4740.2/4711.4; [\ion{Cl}{3}]nn = [\ion{Cl}{3}] 5517.7/5537.9; [\ion{N}{2}]na = [\ion{N}{2}] ($6548.0+6583.5$)/5754.6; [\ion{O}{2}]na = [\ion{O}{2}] ($3726.0+3728.8$)/($7319.5+7330.2$); [\ion{O}{2}]nn = [\ion{O}{2}] 3726.0/3728.8; [\ion{O}{3}]na = [\ion{O}{3}] ($4958.9+5006.8$)/4363.2; and [\ion{S}{2}]nn = [\ion{S}{2}] 6716.4/6730.8.}
\tablenotetext{a}{[\ion{O}{3}]na = [\ion{O}{3}] 4958.9/4363.2.}
\end{deluxetable}
 
\clearpage

\begin{deluxetable}{lc|cc|cc|cc}
\tablecolumns{8}
\tabletypesize{\footnotesize}
\tablewidth{0pc} 
\tablecaption{Cloudy Predicted vs. Observed Diagnostics and Line Intensities}
\tablehead{
\colhead{} & \colhead{} & \multicolumn{2}{|c|}{NGC 6886} & \multicolumn{2}{|c|}{NGC 7027} & \multicolumn{2}{|c}{NGC 7662} \\ \cline{3-4} \cline{5-6} \cline{7-8}
\multicolumn{1}{c}{Ion} & \multicolumn{1}{c}{$\lambda$} & \multicolumn{1}{|c}{Pred.} & \multicolumn{1}{c|}{Obs.} & \multicolumn{1}{|c}{Pred.} & \multicolumn{1}{c|}{Obs.} & \multicolumn{1}{|c}{Pred.} & \multicolumn{1}{c}{Obs.}}
\startdata
$[$\ion{Ar}{3}$]$na & \nodata & 1.17E+02 & 1.09E+02 & 1.08E+02 & 9.79E+01 & 1.06E+02 & 9.68E+01 \\
$[$\ion{Ar}{4}$]$nn & \nodata & 1.53E+00 & 1.29E+00 & 4.13E+00 & 3.70E+00 & 1.07E+00 & 1.26E+00 \\
$[$\ion{Cl}{3}$]$nn & \nodata & 6.27E$-01$ & 3.94E$-01$ & 2.97E$-01$ & 2.90E$-01$ & 8.94E$-01$ & 9.87E$-01$ \\
$[$\ion{N}{2}$]$na & \nodata & 5.40E+01 & 6.38E+01 & 2.26E+01 & 2.20E+01 & 5.30E+01 & 5.05E+01 \\
$[$\ion{O}{2}$]$na & \nodata & 5.73E+00 & 4.50E+00 & 1.06E+00 & 8.70E$-01$ & 1.05E+01 & 8.63E+00 \\
$[$\ion{O}{2}$]$nn & \nodata & 2.97E+00 & 1.53E+00 & 3.65E+00 & 2.76E+00 & 2.33E+00 & 1.70E+00 \\
$[$\ion{O}{3}$]$na & \nodata & 1.23E+02 & 1.22E+02 & 9.23E+01 & 8.75E+01 & 2.84E+01\tablenotemark{a} & 2.33E+01\tablenotemark{a} \\
$[$\ion{S}{2}$]$nn & \nodata & 5.10E$-01$ & 4.84E$-01$ & 4.47E$-01$ & 4.40E$-01$ & 6.09E$-01$ & 6.58E$-01$ \\
\enddata
\tablecomments{Comparison of predicted and observed line intensities relative to H$\beta$ (on the scale $I$(H$\beta)=100$) and diagnostic ratios used in the spectral fits.  Wavelengths are in air for $\lambda>3000$~\AA\ and vacuum otherwise.  Listed wavelengths are in \AA\ except where noted.  The diagnostic ratios are defined as follows: [\ion{Ar}{3}]na = [\ion{Ar}{3}] ($7135.8+7751.1$)/5191.8 ; [\ion{Ar}{4}]nn = [\ion{Ar}{4}] 4740.2/4711.4; [\ion{Cl}{3}]nn = [\ion{Cl}{3}] 5517.7/5537.9; [\ion{N}{2}]na = [\ion{N}{2}] ($6548.0+6583.5$)/5754.6; [\ion{O}{2}]na = [\ion{O}{2}] ($3726.0+3728.8$)/($7319.5+7330.2$); [\ion{O}{2}]nn = [\ion{O}{2}] 3726.0/3728.8; [\ion{O}{3}]na = [\ion{O}{3}] ($4958.9+5006.8$)/4363.2; and [\ion{S}{2}]nn = [\ion{S}{2}] 6716.4/6730.8.}
\tablenotetext{a}{[\ion{O}{3}]na = [\ion{O}{3}] 4958.9/4363.2.}
\end{deluxetable}
 
\clearpage

\begin{deluxetable}{lcccccccccc}
\tablecolumns{11}
\tabletypesize{\footnotesize}
\tablewidth{0pc} 
\tablecaption{Se and Kr Abundance Uncertainties from Various Sources}
\tablehead{
\colhead{PN} & \colhead{$12+$} & \multicolumn{4}{c}{Se Abundance Errors} & \colhead{$12+$} & \multicolumn{4}{c}{Kr Abundance Errors} \\ \cline{3-6} \cline{8-11}
\colhead{Name} & \colhead{Log(Se/H)} & \colhead{$I$} & \colhead{$T_{\rm eff}$} & \colhead{$n_{\rm H}$} & \colhead{Total Err.} & \colhead{Log(Kr/H)} & \colhead{$I$} & \colhead{$T_{\rm eff}$} & \colhead{$n_{\rm H}$} & \colhead{Total Err.}}
\startdata
IC 418\tablenotemark{a} & \nodata & \nodata & \nodata & \nodata & \nodata & 3.90 & 0.144 & 0.17 & 0.03 & 0.21 \\
IC 418\tablenotemark{b} & \nodata & \nodata & \nodata & \nodata & \nodata & 3.77 & 0.179 & 0.14	& 0.10 & 0.23 \\
IC 5117 & 3.67 & 0.044 & 0.02 & 0.02 & 0.05 & 3.76 & 0.111 & 0.01 & 0.01 & 0.11 \\
NGC 6572 & 3.11	& 0.025	& 0.01 & 0.077 & 0.08 & 3.54 & 0.175 & 0.01 & 0.06 & 0.18 \\
NGC 6741 & 3.39	& 0.096	& 0.079	& 0.01 & 0.12 & 3.75 & 0.129 & 0.02 & 0.01 & 0.13 \\
NGC 6790 & 3.10	& 0.044	& 0.015	& 0.015 & 0.05 & 3.24 & 0.134 & 0.01 & 0.05 & 0.14 \\
NGC 6826 & 2.89 & 0.109	& 0.06 & 0.05 & 0.13 & 3.57 & 0.134 & 0.109 & 0.123 & 0.20 \\
NGC 6884 & 3.45 & 0.06 & 0.015 & 0.086 & 0.10 & 3.65 & 0.122 & 0.01 & 0.12 & 0.16 \\
NGC 6886 & 3.59 & 0.045	& 0.07 & 0.079 & 0.11 & 3.81 & 0.16 & 0.088 & 0.05 & 0.18 \\
NGC 7027 & 3.78	& 0.028	& 0.119	& 0.01 & 0.12 & 4.26 & 0.113 & 0.057 & 0.02 & 0.12 \\
NGC 7662 & 3.50	& 0.087	& 0.061	& 0.01 & 0.10 & 3.84 & 0.134 & 0.02 & 0.046 & 0.14 \\
\tableline
\enddata
\tablecomments{1-$\sigma$ abundance uncertainties from various sources are given, along with the total error estimate.  All uncertainties are in dex.}
\tablenotetext{a}{IC 418 model with Rauch NLTE model atmosphere and no mass loss.}
\tablenotetext{b}{IC 418 model with stellar atmosphere with mass-loss (see text, \S3.1.1).}
\end{deluxetable}

\clearpage

\begin{deluxetable}{lcccc}
\tablecolumns{5}
\tablewidth{0pc} 
\tablecaption{Line Intensity Errors from Atomic Data Uncertainties}
\tablehead{
\colhead{PN} & \colhead{$I$([\ion{Se}{4}])} &\colhead{$I$([\ion{Kr}{3}])} & \colhead{$I$([\ion{Kr}{4}])} & \colhead{$I$([\ion{Kr}{5}])} \\
\colhead{Name} & \colhead{2.287~$\mu$m} & \colhead{$\lambda\lambda$6826.7, 21990} & \colhead{$\lambda\lambda$5346.0, 5867.7} & \colhead{$\lambda$6256.1}}
\startdata
\cline{1-5}
\multicolumn{5}{c}{\textit{CT Rate Coefficient Varied}} \\
\cline{1-5}
IC 418 & 0.047 & 0.031 & 0.044 & \nodata \\
NGC 6741 & 0.106 & 0.125 & 0.075 & 0.057 \\
NGC 6884 & 0.074 & 0.152 & 0.052 & 0.036 \\
NGC 7027 & 0.095 & 0.147 & 0.071 & 0.026 \\
\cline{1-5}
\multicolumn{5}{c}{\textit{Low-\emph{T} DR Rate Coefficient Varied}} \\
\cline{1-5}
IC 418 & 0.196 & 0.025 & 0.204 & \nodata \\
NGC 6741 & 0.095 & 0.061 & 0.070 & 0.114 \\
NGC 6884 & 0.081 & 0.109 & 0.053 & 0.120 \\
NGC 7027 & 0.080 & 0.088 & 0.066 & 0.137 \\
\cline{1-5}
\multicolumn{5}{c}{\textit{PI Cross-Section Varied}} \\
\cline{1-5}
IC 418 & 0.215 & 0.090 & 0.248 & \nodata \\
NGC 6741 & 0.126 & 0.136 & 0.091 & 0.120 \\
NGC 6884 & 0.132 & 0.207 & 0.089 & 0.217 \\
NGC 7027 & 0.122 & 0.177 & 0.096 & 0.156 \\
\cline{1-5}
\multicolumn{5}{c}{\textit{RR Rate Coefficient Varied}} \\
\cline{1-5}
IC 418 & 0.067 & 0.045 & 0.120 & \nodata \\
NGC 6741 & 0.030 & 0.021 & 0.040 & 0.089 \\
NGC 6884 & 0.023 & 0.058 & 0.018 & 0.102 \\
NGC 7027 & 0.034 & 0.032 & 0.050 & 0.118 \\
\cline{1-5}
\multicolumn{5}{c}{\textit{Cumulative Uncertainties from Atomic Data Errors}} \\
\cline{1-5}
IC 418 & 0.274 & 0.104 & 0.308 & \nodata \\
NGC 6741 & 0.177 & 0.182 & 0.134 & 0.180 \\
NGC 6884 & 0.162 & 0.257 & 0.112 & 0.249 \\
NGC 7027 & 0.165 & 0.227 & 0.136 & 0.218 \\
\tableline
\enddata
\tablecomments{Predicted line intensity errors (in dex) from atomic data uncertainties.  See \S3.1.3.}
\end{deluxetable}

\clearpage

\begin{deluxetable}{cccccccccc}
\tablecolumns{10}
\tablewidth{0pc} 
\tablecaption{Values of Physical Parameters in Grid of Cloudy models}
\tablehead{\multicolumn{10}{c}{$T_{\rm eff}$ (10$^3$ K)}}
\startdata
50 & 55 & 60 & 65 & 70 & 75 & 80 & 85 & 90 & 95 \\
100 & 110 & 120 & 130 & 140 & 150 & 160 & 170 & 180 & \\
\tableline
\multicolumn{10}{c}{$n_{\rm H}$ (10$^3$ cm$^{-3}$)} \\
\tableline
1 & 2 & 3 & 4 & 5 & 6 & 7 & 8 & 9 & 10 \\
12.5 & 15 & 20 & 25 & 30 & 40 & 50 & 60 & 70 & 80 \\
 & & & & 90 & 100 & & & & \\
\tableline
\multicolumn{10}{c}{log($U$)} \\
\tableline
$-3.0$ & $-2.0$ & $-1.5$ & $-1.0$ & $-0.5$ & 0.0 & 0.5 & 1.0 & 2.0 & \\
\tableline
\enddata
\tablecomments{Values of the input parameters varied in our grid of Cloudy models are listed.  All combinations of the listed parameters were input into our grid, for a total of 3761 models.}
\end{deluxetable}

\clearpage

\begin{deluxetable}{lcccc}
\tablecolumns{5}
\tablewidth{0pc} 
\tablecaption{Comparison of ICF- and Model-Derived Kr Abundances}
\tablehead{
\colhead{PN} & \colhead{$n$(Kr$^{++}$)/} & \colhead{} & \colhead{$12+$ Log(Kr/H)} & \colhead{$12+$ Log(Kr/H)} \\
\colhead{Name} & \colhead{$n$(H$^+$)$\times 10^9$} & \colhead{ICF(Kr)\tablenotemark{a}} & \colhead{ICF} & \colhead{Model}}
\startdata
IC 418 & 3.18$\pm$0.79 & 2.05$\pm$1.36 & 3.81$\pm$0.33 & 3.90$\pm$0.21\\
IC 5117 & 1.36$\pm$0.37 & 4.75$\pm$1.96 & 3.81$\pm$0.24 & 3.76$\pm$0.11 \\
NGC 6572 & 1.16$\pm$0.30 & 3.10$\pm$2.13 & 3.56$\pm$0.40 & 3.54$\pm$0.18 \\
NGC 6741 & 1.80$\pm$0.29 & 4.74$\pm$2.42 & 3.93$\pm$0.26 & 3.75$\pm$0.13 \\
NGC 6790 & 0.24$\pm$0.13 & 5.08$\pm$2.45 & 3.08$\pm$0.40 & 3.24$\pm$0.14 \\
NGC 6826 & 1.03$\pm$0.44 & 2.16$\pm$1.79 & 3.35$\pm$0.48 & 3.57$\pm$0.20 \\
NGC 6884 & 0.62$\pm$0.23 & 5.35$\pm$2.32 & 3.52$\pm$0.30 & 3.65$\pm$0.16 \\
NGC 6886 & 1.67$\pm$0.37 & 5.33$\pm$2.51 & 3.95$\pm$0.25 & 3.81$\pm$0.18 \\
NGC 7027 & 2.42$\pm$0.46 & 6.13$\pm$2.69 & 4.17$\pm$0.23 & 4.26$\pm$0.12 \\
NGC 7662 & 0.40$\pm$0.14 & 7.98$\pm$3.14 & 3.50$\pm$0.25 & 3.84$\pm$0.14 \\
\tableline
\enddata
\tablecomments{Comparison of Kr abundances derived from ICFs and from our Cloudy models.  The solar Kr abundance is $12$~+~log(Kr/H)~=~$3.28\pm0.08$ (Asplund et al.\ 2005).}
\tablenotetext{a}{Equation (1) is used to determine ICF(Kr) for all objects except IC~418, where Equation (2) is used (due to the small Ar$^{++}$/Ar fraction and hence large uncertainty in the corresponding ICF).}
\end{deluxetable}

\clearpage

\begin{deluxetable}{lcccc}
\tablecolumns{5}
\tablewidth{0pc} 
\tablecaption{Comparison of ICF- and Model-Derived Se Abundances}
\tablehead{
\colhead{PN} & \colhead{$n$(Se$^{3+}$)/} & \colhead{} & \colhead{$12+$ Log(Se/H)} & \colhead{$12+$ Log(Se/H)} \\
\colhead{Name} & \colhead{$n$(H$^+$)$\times 10^9$} & \colhead{ICF(Se)} & \colhead{ICF} & \colhead{Model}}
\startdata
IC 418 & \nodata & 22.26$\pm$8.56 & \nodata & \nodata \\
IC 5117 & 2.20$\pm$0.24 & 2.32$\pm$0.96 & 3.71$\pm$0.20 & 3.67$\pm$0.05 \\
NGC 6572 & 0.69$\pm$0.05 & 2.32$\pm$0.71 & 3.20$\pm$0.14 & 3.11$\pm$0.08 \\
NGC 6741 & 0.59$\pm$0.15 & 8.12$\pm$3.62 & 3.68$\pm$0.25 & 3.39$\pm$0.12 \\
NGC 6790 & 0.56$\pm$0.06 & 2.50$\pm$0.73 & 3.15$\pm$0.14 & 3.10$\pm$0.05 \\
NGC 6826 & 0.47$\pm$0.15 & 2.14$\pm$0.97 & 3.00$\pm$0.27 & 2.89$\pm$0.13 \\
NGC 6884 & 1.43$\pm$0.22 & 2.12$\pm$0.50 & 3.48$\pm$0.13 & 3.45$\pm$0.10 \\
NGC 6886 & 0.93$\pm$0.11 & 3.43$\pm$2.03 & 3.51$\pm$0.30 & 3.59$\pm$0.11 \\
NGC 7027 & 1.51$\pm$0.11 & 3.23$\pm$1.93 & 3.69$\pm$0.30 & 3.78$\pm$0.12 \\
NGC 7662 & 1.43$\pm$0.32 & 4.64$\pm$2.47 & 3.82$\pm$0.29 & 3.50$\pm$0.10 \\
\tableline
\enddata
\tablecomments{Comparison of Se abundances derived from ICFs and from our Cloudy models.  The solar Se abundance is $12$~+~log(Se/H)~=~$3.33\pm0.04$ (Asplund et al.\ 2005).}
\end{deluxetable}

\clearpage

\begin{deluxetable}{lccl}
\tablecolumns{4}
\tablenum{A1}
\tablewidth{0pc} 
\tablecaption{Possible Identifications for the $\lambda$5346.0 Feature}
\tablehead{
\colhead{} & \colhead{Wavelength} & \colhead{$E_{\rm U}$--$E_{\rm L}$} & \colhead{Other Multiplet} \\
\colhead{Ion} & \colhead{(\AA)} & \colhead{(eV)} & \colhead{Lines (\AA)}}
\startdata
$[$\ion{Kr}{4}$]$ & 5346.0 & 2.318--0.0 & 5867.7 \\
\ion{C}{1} & 5345.6 & 11.165--8.847 & 5350.6, 5356.5 \\
\ion{S}{2} & 5345.7 & 17.386--15.068 & 5320.7, 5346.1 \\
\ion{Fe}{2} & 5345.7 & 12.786--10.467 & 5323.7, 5351.5 \\
\ion{C}{3} & 5345.9 & 42.170--39.852 & 5337.4, 5341.5, 5348.8 \\
 &  &  & 5353.2, 5359.9 \\
\ion{N}{2} & 5345.9 & 29.237--26.919 & 5344.4, 5347.4, 5371.5 \\
 &  &  & 5374.6, 5388.6, 5416.2 \\
\ion{Fe}{2} & 5346.0 & 12.818--10.500 & 5353.4, 5401.1, 5407.2 \\
 &  &  & 5459.6 \\
\ion{S}{2} & 5346.1 & 17.386--15.068 & 5320.7, 5346.1 \\
\ion{Fe}{3} & 5346.4 & 25.417--23.099 & 5314.6, 5322.2, 5335.9 \\
 &  &  & 5338.0, 5343.7, 5354.9 \\
 &  &  & 5364.9, 5365.3, 5369.6 \\
 &  &  & 5377.0 \\
\ion{N}{1} & 5346.5 & 14.314--11.996 & 5310.4, 5333.4 \\
\tableline
\enddata
\tablecomments{Potential identifications of $\lambda$5346.0 are given, with the upper and lower energy levels of the transition and wavelengths of other multiplet members.}
\end{deluxetable}

\clearpage

\begin{deluxetable}{llll}
\tablecolumns{4}
\tablewidth{0pc} 
\tablenum{B1}
\tablecaption{Transition Probability and Effective Collision Strength References}
\tablehead{
\colhead{Ion} & \colhead{$A_{\rm ij}$} & \colhead{$\Omega_{\rm ij}$}}
\startdata
\ion{Se}{4} & Bi\'{e}mont \& Hansen (1987) & Butler (2007)\tablenotemark{a}\\
\ion{Kr}{3} & Bi\'{e}mont \& Hansen (1986b) & Sch\"{o}ning (1997) \\
\ion{Kr}{4} & Bi\'{e}mont \& Hansen (1986a) & Sch\"{o}ning (1997) \\
\ion{Kr}{5} & Bi\'{e}mont \& Hansen (1986a) & Sch\"{o}ning (1997) \\
\tableline
\enddata
\tablenotetext{a}{In preparation.  The effective collision strength of [\ion{Se}{4}]~2.287~$\mu$m at $T_{\rm e}=10^4$~K is 7.66.}
\end{deluxetable}

\clearpage

\begin{deluxetable}{ccllc}
\tablecolumns{5}
\tablewidth{0pc} 
\tablenum{B2}
\tablecaption{Adopted Threshold Photoionization Cross-Sections and Radiative Recombination Fit Parameters}
\tablehead{
\colhead{} & \colhead{} & \colhead{IP} & \colhead{$\sigma_{\rm th}$} & \colhead{$A_{\rm rr}$} \\
\colhead{Element} & \colhead{Charge} & \colhead{(eV)} & \colhead{(Mb)} & \colhead{(10$^{-11}$ cm$^3$ s$^{-1}$)}}
\startdata
Se & 0 &    9.752E+00 &    1.100E+01 &    3.120E$-01$ \\
  & 1 &    2.119E+01 &    3.810E+00 &    3.360E$-01$ \\
  & 2 &    3.082E+01 &    1.740E+00 &    4.830E$-01$ \\
  & 3 &    4.294E+01 &    6.260E$-01$ &    6.430E$-01$ \\
  & 4 &    6.832E+01 &    7.870E$-01$ &    1.220E+00 \\
  & 5 &    8.170E+01 &    3.290E$-01$ &    1.620E+00 \\
  & 6 &    1.554E+02 &    2.310E+00 &    2.380E+00 \\
  & 7 &    1.911E+02 &    1.690E+00 &    3.670E+00 \\
\enddata
\tablecomments{Ionization potentials, threshold photoionization cross-sections $\sigma_{\rm th}$ in megabarns (10$^{-18}$ cm$^2$), and RR rate coefficients $A_{\rm rr}$ are listed.  These are used in Equations~(B2) and (B3) to determine the photoionization cross-sections and radiative rate coefficients as functions of photon energy and electron temperature, respectively.}
\tablenotetext{a}{$\sigma_{\rm th}$ of \ion{Kr}{3} from equation (B1) multiplied by 2.  See \S3.1.}
\tablenotetext{b}{$\sigma_{\rm th}$ of \ion{Kr}{4} from equation (B1) multiplied by 0.25.  See \S3.1.}
\end{deluxetable}

\clearpage

\begin{deluxetable}{cccccccccc}
\tablecolumns{9}
\tabletypesize{\scriptsize}
\tablenum{B3}
\tablewidth{0pc} 
\tablecaption{Adopted High-Temperature Dielectronic Recombination and Collisional Ionization Fit Parameters}
\tablehead{
\colhead{} & \colhead{} & \colhead{$A_{\rm di}$} & \colhead{$T_0$} & \colhead{$B_{\rm di}$} & \colhead{$T_1$} & \colhead{$P$} & \colhead{$A_{\rm ci}$} & \colhead{$X$} & \colhead{$K$}\\ 
\colhead{Element} & \colhead{Charge} & \colhead{(cm$^3$ s$^{-1}$ K$^{1.5}$)} & \colhead{(K)} & \colhead{} & \colhead{(K)} & \colhead{} & \colhead{(cm$^3$ s$^{-1}$)} & \colhead{} & \colhead{}}
\startdata
Kr & 0 & 3.612E$-09$ & 1.806E+04 & 1.317E$-10$ & 2.948E+05 & 0 & 7.761E$-10$ & 1.169E$-05$ & 4.433E$-01$ \\
  & 1 & 9.360E$-09$ & 2.894E+04 & 1.129E$-01$ & 1.052E+04 & 0 & 3.260E$-10$ & 9.500E$-06$ & 4.599E$-01$ \\
  & 2 & 1.555E$-08$ & 3.926E+04 & 1.626E$-01$ & 3.345E+04 & 0 & 1.370E$-10$ & 1.121E$-05$ & 4.623E$-01$ \\
  & 3 & 1.798E$-08$ & 4.497E+04 & 2.608E$-01$ & 8.352E+04 & 0 & 5.983E$-11$ & 1.142E$-05$ & 4.567E$-01$ \\
  & 4 & 1.560E$-08$ & 4.589E+04 & 4.777E$-01$ & 1.483E+05 & 0 & 2.326E$-11$ & 1.186E$-05$ & 4.356E$-01$ \\
  & 5 & 7.783E$-09$ & 4.048E+04 & 1.412E+00 & 2.487E+05 & 0 & 9.163E$-12$ & 7.076E$-06$ & 3.378E$-01$ \\
  & 6 & 5.494E$-09$ & 5.234E+04 & 2.780E+00 & 3.663E+05 & 0 & 5.930E$-12$ & 9.025E$-06$ & 3.121E$-01$ \\
  & 7 & 2.018E$-09$ & 1.198E+06 & 4.793E+01 & 1.556E+02 & 0 & 4.807E$-12$ & 8.671E$-06$ & 3.031E$-01$ \\
\enddata
\tablecomments{Fit parameters for high-temperature DR and CI, used in equations (B6) and (B7), respectively, are shown.  Fits were performed to the DR data of Fournier et al.\ (2000) and the CI data of Loch et al.\ (2002).}
\end{deluxetable}

\clearpage

\begin{figure}
\epsscale{0.8}
\plotone{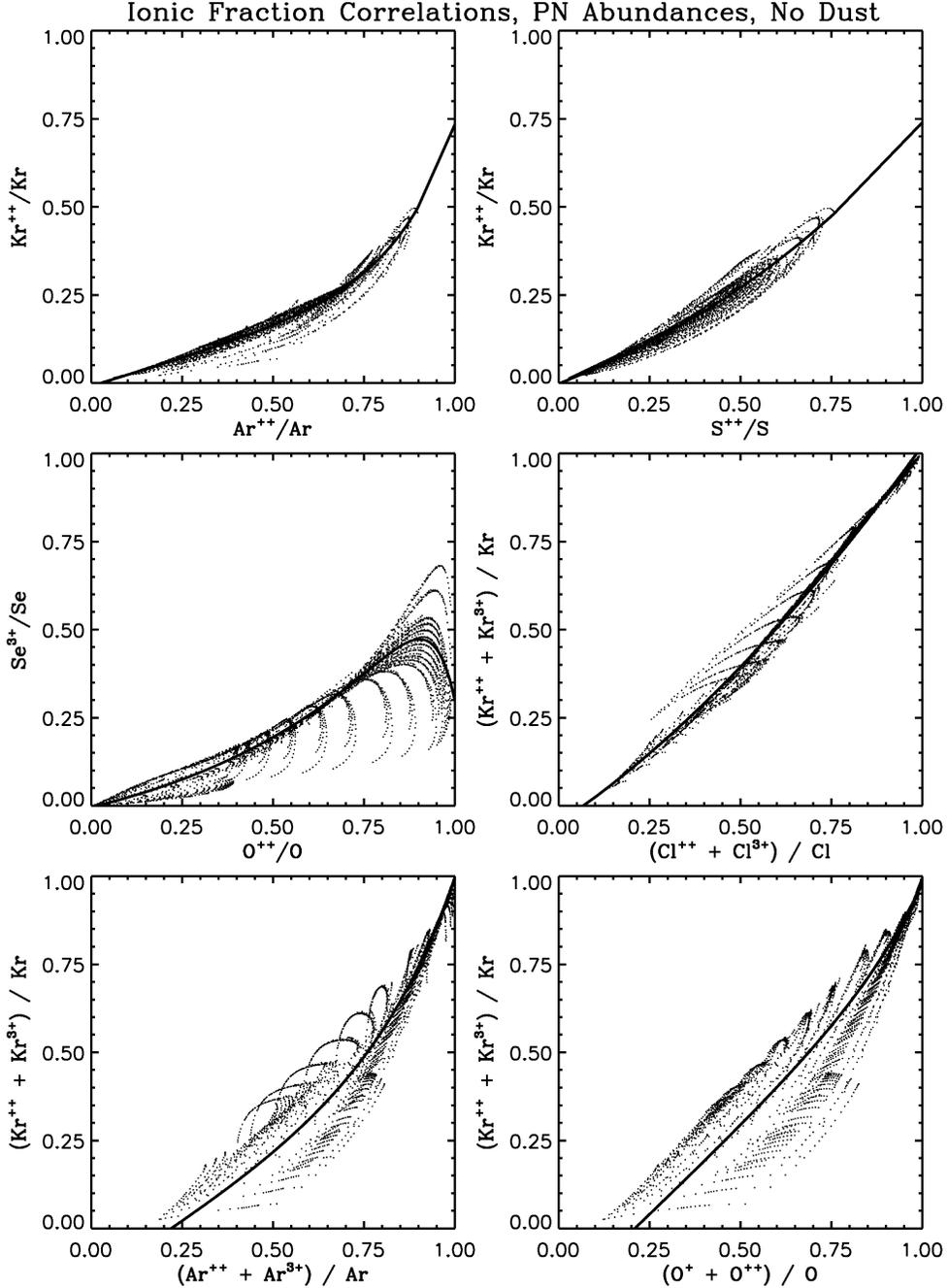}
\figcaption[f1.eps]{Correlations between Kr and Se ionic fractions and those of commonly detected elements are plotted.  Each dot corresponds to a single model from our grid of Cloudy models.  The default PN abundances of Cloudy are assumed, and dust grains are not included.  In some of the panels, families of curves are seen; these generally correspond to single values of $T_{\rm eff}$, with each curve produced by variations in $n_{\rm H}$ and $U$.  Fits to these correlations (\S4) are overplotted as thick solid lines.}
\end{figure}

\clearpage

\begin{figure}
\epsscale{0.8}
\plotone{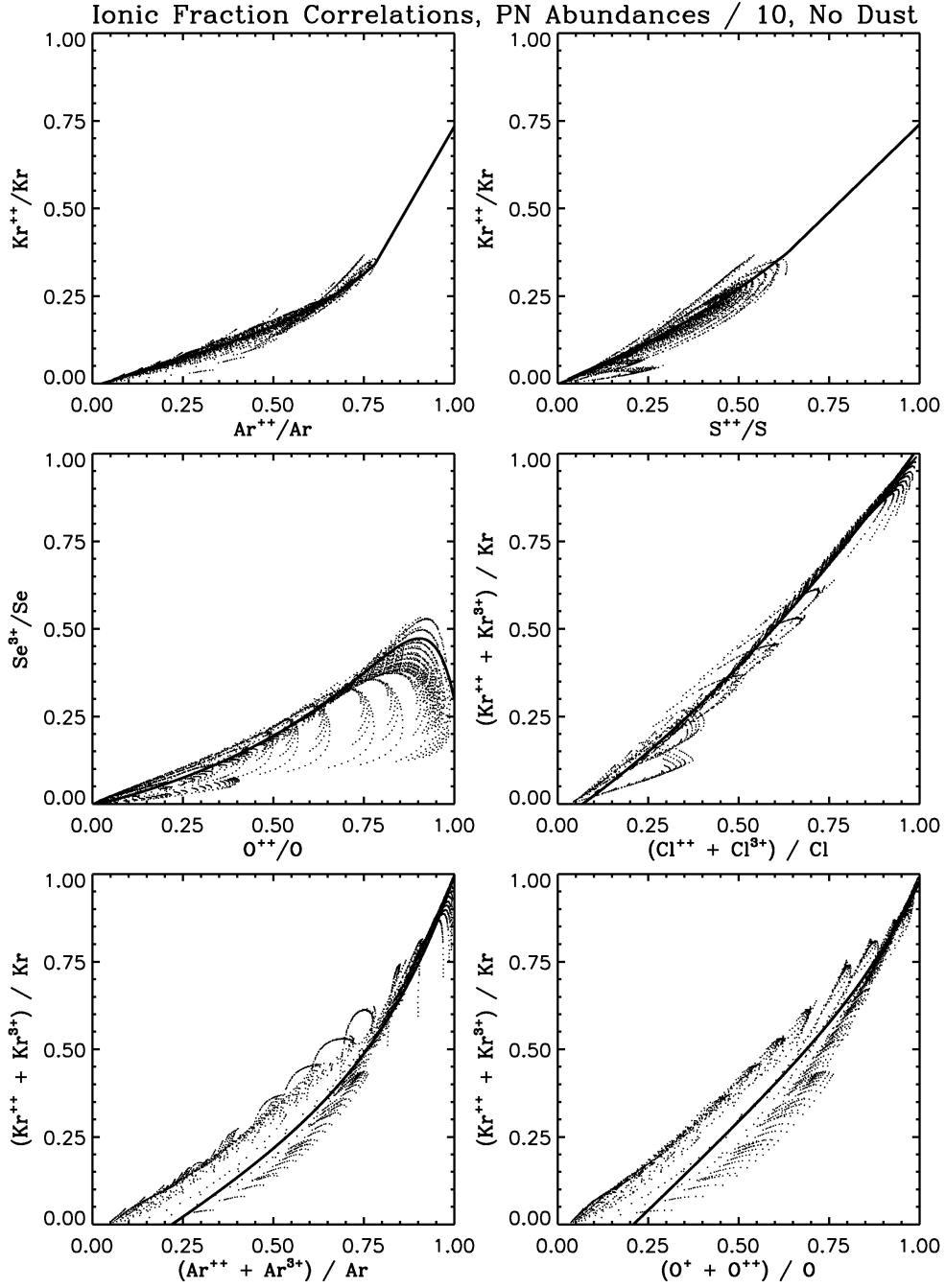}
\figcaption[f2.eps]{Same as Figure~1, except the default PN abundances of Cloudy are divided by ten for all elements heavier than He, and NLTE stellar atmospheres with halo abundances are used.  This shows that our Se and Kr ICFs are minimally affected by overall (lower) metallicity, with the exception of the correlations shown in the two bottom panels.  The fits shown are the same as in Figure~1.}
\end{figure}

\clearpage

\begin{figure}
\epsscale{0.8}
\plotone{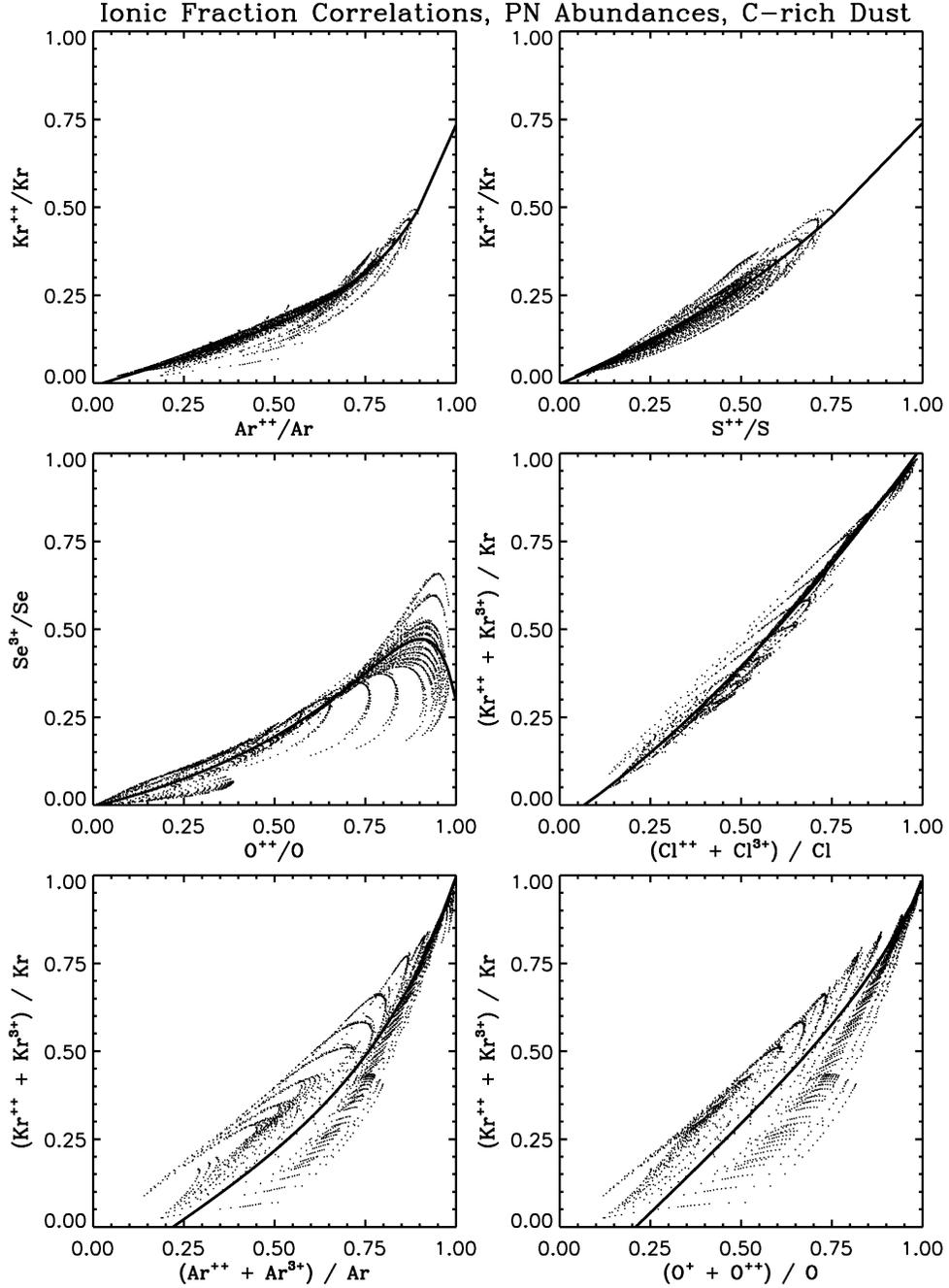}
\figcaption[f3.eps]{Same as Figure~1, except C-rich dust is included in the models.  The fits shown are the same as in Figure~1.  The inclusion of dust physics produces negligible effects on the ionic fraction correlations.  The same conclusion was found for PNe with O-rich dust (not plotted).}
\end{figure}

\end{document}